%% file: main.tex
\documentclass[11pt,british]{article}
\usepackage[utf8]{inputenc}
\usepackage{mathtools}
\usepackage{amsmath}
\usepackage{amsthm}
\usepackage{amssymb}
\usepackage{setspace}
\usepackage{microtype}
\usepackage{antmath}
\usepackage{babel}
\usepackage{stmaryrd}
\usepackage{slashed}
\usepackage{cancel}
\usepackage{bbm}
\makeatletter
\usepackage{tikz}
\usepackage{tikz-cd}
\usetikzlibrary{backgrounds}
\usepackage[framemethod=tikz]{mdframed}
\AtBeginEnvironment{mdframed}{%
\tikzset{every picture/.style={}}%
}
\mdfsetup{roundcorner=.5ex}

\theoremstyle{definition}
\newtheorem*{defn*}{Definition}

\usepackage{jheppub}                   
\usepackage[linecolor=blue,backgroundcolor=blue!25,bordercolor=blue,textsize=scriptsize]{todonotes}

\makeatletter
\gdef\@fpheader{\ }                    
\makeatother

\usepackage{setspace}
\usepackage{slashed}	 	
\usepackage{amsfonts} 		
\SetTracking{encoding={*}, shape=sc}{0} 	
\usepackage{color} 		
\definecolor{darkblue}{rgb}{0.0,0.0,0.3} 	
\allowdisplaybreaks		
\date{\today} 		
\numberwithin{equation}{section}	

\makeatletter
\g@addto@macro\bfseries{\boldmath}
\makeatother

\let\originalleft\left
\let\originalright\right
\renewcommand{\left}{\mathopen{}\mathclose\bgroup\originalleft}
\renewcommand{\right}{\aftergroup\egroup\originalright}

\newcommand{\Q}{C}

\newcommand{\QQ}{\mathcal{C}}

\newcommand{\adQ}{\mathfrak{C}}

\title{Heterotic backgrounds via generalised geometry: moment maps and moduli}

\author[a]{Anthony Ashmore,}
\emailAdd{aashmore@sas.upenn.edu}
\author[b]{Charles Strickland-Constable,}
\emailAdd{c.strickland-constable@herts.ac.uk}
\author[c]{David Tennyson}
\emailAdd{d.tennyson16@imperial.ac.uk}
\author[c]{\\and Daniel Waldram}
\emailAdd{d.waldram@imperial.ac.uk}

\affiliation[a]{Department of Physics and Astronomy, University of Pennsylvania,\\
Philadelphia, PA 19104, USA}
\affiliation[b]{School of Physics, Astronomy and Mathematics, University of Hertfordshire, \\College Lane, Hatfield, AL10 9AB, UK}
\affiliation[c]{Department of Physics,
	Imperial College London, \\
	Prince Consort Road, London, SW7 2AZ, UK} 

\subheader{\hfill\textrm{Imperial/TP/19/DW/2}}

\abstract{We describe the geometry of generic heterotic backgrounds preserving minimal supersymmetry in four dimensions using the language of generalised geometry. They are characterised by an $\SU3\times\Spin{6+n}$ structure within $\Orth{6,6+n}\times\mathbb{R}^+$ generalised geometry. Supersymmetry of the background is encoded in the existence of an involutive subbundle of the generalised tangent bundle and the vanishing of a moment map for the action of diffeomorphisms and gauge symmetries. We give both the superpotential and the Kähler potential for a generic background, showing that the latter defines a natural Hitchin functional for heterotic geometries. Intriguingly, this formulation suggests new connections to geometric invariant theory and an extended notion of stability. Finally we show that the analysis of infinitesimal deformations of these geometric structures naturally reproduces the known cohomologies that count the massless moduli of supersymmetric heterotic backgrounds.}

\begin{document}
\maketitle


\section{Introduction}

The conditions for a generic minimally supersymmetric compactification of the heterotic string to four-dimensional Minkowski space were first given by Hull and Strominger in the 1980s~\cite{Hull86d,Strominger86}. Since then there has been significant progress in understanding the properties of these compactifications from both worldsheet~\cite{AEL08, Kreuzer:2010ph, McOrist:2010ae, Beccaria:2010yp, McOrist:2011bn, MS11, Blaszczyk:2011ib, Quigley:2011pv, Nibbelink:2012wb} and spacetime perspectives~\cite{BBD+03b, BBG+04, BT06, CLP12, Lust:1986ix, BR89b, FI01, Becker:2002sx, GMW03, LopesCardoso2003sp, LCD+03b, Becker:2006et, Garcia-Fernandez:2013gja,li2005,Fu:2006vj, Andreas:2010qh, BS09, Andreas:2010cv}. A classic solution has been to take the internal space to be a Calabi--Yau manifold with a gauge bundle satisfying the hermitian Yang--Mills equations~\cite{CHSW85}. In recent years a great amount of mathematics has been developed to use such constructions to find four-dimensional models with chiral fermions and Standard Model gauge groups~\cite{BCD+12, Braun:2005bw, BHO+05b, Braun:2005nv, Anderson:2010mh, Anderson:2011ns, Anderson:2012yf, Anderson:2011cza}. Such models generically admit a large number of massless modes in four dimensions, spurring much work on ways to stabilise these moduli. 

In the hope of stabilising moduli, there has been significant effort devoted to turning on flux or torsion in heterotic backgrounds~\cite{Anderson:2010mh, Anderson:2011cza, Anderson:2011ty}. Despite this attention, once one moves away from Calabi--Yau examples, heterotic backgrounds and their moduli remain poorly understood.  For a long time there was little progress on understanding even the infinitesimal moduli around generic points in the moduli space, particularly those corresponding to non-Kähler solutions of the Hull--Strominger system. Recently there has been some success with determining the infinitesimal moduli in terms of deformations of a holomorphic structure on a particular Courant algebroid on the internal space~\cite{BT06, Anderson:2010mh, AGL+09, Anderson:2011ty, MS11, OS14b, Anderson:2014xha, OHS16, GRT15, Becker:2006xp, Cyrier:2006pp, McOrist:2019mxh,OS14b}. This has now been extended to finite deformations, where the moduli are controlled by a degree-three $L_{\infty}$-algebra satisfying a set of Maurer--Cartan like equations~\cite{Ashmore:2018ybe}. Results on finite deformations of so-called holomorphic string algebroids have also recently appeared in~\cite{GRT18,GRST18}. There have also been many results regarding the superpotential and Kähler potential for compactifications on Calabi--Yau manifolds~\cite{Anderson:2010mh, Anderson:2011cza}, manifolds with flux and torsion~\cite{Benmachiche:2008ma,OHS16, GLM04, LCD+03b}, and more general non-Kähler manifolds~\cite{BBD+03b, BBG+04, BBD+03}. However there has yet to be a unified approach that describes the generic properties of heterotic backgrounds. It is the aim of this paper to provide such a unified description using generalised geometry. We shall see there is a geometric description of generic $\mathcal{N}=1$, $D=4$ heterotic backgrounds, with objects such as the superpotential and Kähler potential appearing naturally. We will also see how this new approach is well suited to finding moduli. Much of this work builds on \cite{N1paper}, where analogous structures were identified in M-theory and type II theories.

Generalised geometry provides a natural framework for analysing the structure of supersymmetric backgrounds with a Minkowski external space in terms of integrable $G$-structures on a generalised tangent bundle. Here, integrability is defined to mean the existence of a torsion-free generalised connection that is compatible with the geometric structure.\footnote{See e.g.~\cite{AW15,CS16} for definitions of generalised connections, torsion and compatibility.} It is known~\cite{CSW11b} that a generic bosonic field configuration of the NSNS sector of type II theories is characterised by an integrable $\Orth{d}\times \Orth{d}$ structure in $\Orth{d,d}\times\mathbb{R}^{+}$ generalised geometry, where $d$ is the number of internal dimensions. This was extended to the entire bosonic sector of type II or M-theory in \cite{CSW14} and \cite{CSW11}, where the relevant integrable structure is the maximally compact subgroup $H_{d}\subset \Ex{d(d)}\times \mathbb{R}^{+}$. Here $d$ (or $d-1$) is the dimension of the internal space of the M-theory (or type II) background. Supersymmetry of the background further constrains these geometric structures, with preserved supersymmetry being equivalent to the existence of an integrable $G_{d}^\mathcal{N}\subset \tilde{H}_{d}$ structure~\cite{CSW14b,CS16}, where $\tilde{H}_{d}$ is the double cover of $H_{d}$, and $G_{d}^\mathcal{N}$ depends on the dimension of the internal space $d$ (or $d-1$) and the number of preserved supersymmetries $\mathcal{N}$. A description of the relevant $G$-structures for backgrounds preserving eight supercharges in four dimensions was given in \cite{GLSW09}, with a description in terms of invariant generalised tensors for  various dimensions given in \cite{AW15}.

In \cite{N1paper}, we analysed the geometry and moduli of M-theory and type II backgrounds with a four-dimensional Minkowski factor preserving $\mathcal{N}=1$ supersymmetry. These backgrounds are characterised by integrable $\SU7$ structures~\cite{PW08,CSW14b}. Such structures are defined by a nowhere-vanishing generalised tensor $\psi$ transforming in the $\rep{912}_{\rep 3}$ of $\Ex{7(7)}\times\mathbb{R}^{+}$. Important quantities, such as the superpotential and Kähler potential of the effective theory, were given in terms of $\psi$. Interestingly, a slightly weaker $\mathbb{R}^{+}\times \Uni7$ structure can be defined in terms of a certain subbundle of the generalised tangent bundle, similar to an almost complex structure. The integrability of the $\SU7$ structure is then equivalent to the closure of this subbundle under the Courant bracket, plus a moment map condition. Using the equivalence of Kähler quotients and complexified quotients, we argued that the $\SU7$ moduli can be found by considering deformations of the $\mathbb{R}^{+}\times \Uni7$ structure up to complexified generalised diffeomorphisms. We will see that a completely analogous story appears for heterotic backgrounds. In the case of pure $\Orth{d,d}$ generalised geometry, a relation between supersymmetry of the underlying sigma model and integrability of a subbundle has appeared in~\cite{Hull:2018jkr}.

The relevant generalised geometry for heterotic (or type I) backgrounds is $\Orth{6,6+n}\times\mathbb{R}^{+}$~\cite{CMTW14,Ashmore:2018ybe,CSW14b,Garcia-Fernandez:2013gja}, where $n$ is the dimension of the gauge group $G$. As we will discuss, a general $\mathcal{N}=1$, $D=4$ heterotic background is characterised by an integrable $\SU3\times\Spin{6+n}$ structure~\cite{Ashmore:2018ybe}. This structure can be defined by a nowhere-vanishing generalised tensor $\psi$ transforming in the $\rep{220}_{\rep1}$ representation of $\Orth{6,6+n}\times\mathbb{R}^{+}$. Following \cite{N1paper}, we will also define a weaker $\mathbb{R}^{+}\times\Uni3\times \Spin{6+n}$ structure, equivalent to a certain subbundle $L_{-1}$ of the complexified generalised tangent bundle. We will give the conditions on these structures for them to be integrable. We will see that integrability of the $\mathbb{R}^{+}\times\Uni3\times \Spin{6+n}$ structure is given by an involutivity condition on the subbundle $L_{-1}$. (Such an involutivity condition has previously appeared in a generalised geometric analysis of non-linear sigma models whose target spaces are ``strong Kähler with torsion''~\cite{Hull:2019iof}.) Integrability of the full $\SU3\times\Spin{6+n}$ structure, equivalent to supersymmetry for the background, requires an additional condition which takes the form of the vanishing of a moment map for the action of diffeomorphisms and gauge transformations on the space of $\SU3\times\Spin{6+n}$ structures. We will show that these geometric conditions are equivalent to the equations of the Hull--Strominger system\footnote{\textbf{Note added:} Shortly after this paper first appeared, an independent moment-map interpretation of the Hull--Strominger system in terms of holomorphic Courant algebroids was given in~\cite{Garcia-Fernandez:2020awc}.}. Moreover, the split into involutivity and a moment map mirrors the split into F- and D-term conditions in the four-dimensional effective theory.

These structures are completely generic, defined for any value of the fluxes and field strengths in the theory. From these we are able to give expressions for the perturbative Kähler potential and superpotential without resorting to Kaluza--Klein reduction or assuming an expansion in terms of harmonic forms on the internal space. Both of these expressions are written in terms of $\psi$ and hence are covariant. These geometric structures are ideal for investigating the moduli of general heterotic compactifications. In particular, the appearance of a complexified quotient by the symmetries naturally gives rise to a cohomology which counts the moduli. Using this perspective, we recover the cohomology found in \cite{OS14b,Ashmore:2018ybe}.

The paper is structured as followed. In section \ref{sec:Hull-Strominger Review} we give a brief introduction to the Hull--Strominger system as well as a review of what is known about the infinitesimal moduli problem in terms of deformations of a holomorphic Courant algebroid. In section \ref{sec:Gen Geom for heterotic} we describe the formulation of heterotic backgrounds in terms of $\Orth{6,6+n}\times\mathbb{R}^{+}$ generalised geometry, giving definitions for the generalised tensor $\psi$, and the bundle $L_{-1}$. We do this first for the case with no gauge bundle in section \ref{sec:U(3)xSU(4)}, then we reintroduce the gauge bundle in \ref{sec:U(3)xSpin(6+n)}. We also discuss the equivalence between supersymmetry and integrability for the structures. In section \ref{sec:involutivity}, we explore involutivity of $L_{-1}$ and give the superpotential in terms of $\psi$. We also show how these are related to the F-term conditions of the Hull--Strominger system. In section \ref{sec:Kahler, moment and D}, we give the Kähler potential on the space of structures and derive a moment map for the action of generalised diffeomorphisms. We compute both of these explicitly and show that the moment map reproduces the final supersymmetry conditions, now with a geometric interpretation. This reinterpretation of the supersymmetry conditions as the vanishing of some moment map provides some interesting links with geometric invariant theory which we highlight in section \ref{sec:GIT}. In section \ref{sec:moduli} we find the infinitesimal moduli and show that they are related to the previously known $\bar D$ cohomology. We finish with some general comments and further directions in section \ref{sec:conclusions}. Appendix \ref{ap:O(6,6+n) generalised geometry} contains our conventions and appendix \ref{sec:explicit_moment} contains detailed calculations of the superpotential, the Kähler potential, and the moment map.

\section{Review of the Hull--Strominger system}\label{sec:Hull-Strominger Review}

We begin with a  review of the Hull--Strominger system~\cite{Hull86,Strominger86}. This is a set of equations describing the geometry of general $\mathcal{N}=1$ backgrounds of the heterotic string on a ten-dimensional manifold $M$ that is a product of a six-dimensional manifold $X$ with four-dimensional Minkowski space $M=\mathbb{R}^{3,1}\times X$, with trivial warp factor in the string frame. 

The condition of $\mathcal{N}=1$ supersymmetry implies the existence of a global nowhere-vanishing spinor $\epsilon$ on $X$. This defines an $\SU3$ structure on $X$ which can be equivalently described in terms of a complex three-form $\Omega$ (with a compatible almost complex structure $I$) and a real two-form $\omega$ satisfying
\begin{equation}
\Omega\wedge\omega = 0, \qquad \tfrac{1}{8}\ii\,\Omega\wedge\bar{\Omega} = \tfrac{1}{6} \omega\wedge\omega\wedge\omega.
\end{equation}
As usual, the forms are defined as bilinears in the spinor $\epsilon$
\begin{equation}
\Omega_{mnp} = \epsilon^{\text{T}}\gamma_{mnp}\epsilon ,
\qquad 
\omega_{mn} = -\ii \,\epsilon^{\dagger}\gamma_{mn}\epsilon . 
\end{equation}
The supersymmetry conditions in the form of the Killing spinor equations imply that this $\SU3$ structure is not integrable but instead satisfies 
\begin{equation}
\label{eq:conformally holomorphic-balanced} 
\dd(\ee^{-2\varphi}\Omega) = 0 ,  \qquad
\dd(\ee^{-2\varphi}\omega\wedge\omega) = 0 , 
\end{equation}
where $\varphi$ is the dilaton. These conditions are known as ``conformally holomorphic'' and ``conformally balanced'' respectively. Note that the first condition implies that $X$ has an integrable complex structure whose canonical bundle is holomorphically trivial. 

Heterotic compactifications come with a connection $A$ on a vector bundle $V \to X$ whose field strength $F$ is valued in $\End(V)$, and a connection $\Theta$ on the tangent bundle $T$ whose curvature $R$ is valued in $\End(T)$. Supersymmetry implies that both connections are instantons~\cite{Ivanov10,MS11b}
\begin{equation}
F_{0,2}= 0,\qquad \omega^{\sharp}\lrcorner F =0, \qquad\text{and}\qquad R_{0,2} = 0,\qquad   \omega^{\sharp}\lrcorner R=0,
\end{equation}
where $\omega^\sharp$ is $\omega$ with its indices raised using the metric on $X$ and a subscript indicates the $(0,2)$-form part of the curvature with respect to the complex structure defined by the $\SU{3}$ structure. In other words, $V$ and $T$ must be holomorphic vector bundles with connections that solve the hermitian Yang--Mills equations with zero slope. A theorem due to Donaldson--Uhlenbeck--Yau then guarantees a unique solution provided $V$ and $T$ are polystable~\cite{Donaldson85,UY86}. 

The final supersymmetry condition is the anomaly cancellation condition. This couples the intrinsic torsion of the $\SU3$ structure with the $B$ field and the connections. It is given by
\begin{equation}
\ii(\partial-\bar{\partial})\omega = H \coloneqq \dd B + \tfrac{1}{4}\alpha'\left(\omega_{3}(A)-\omega_{3}(\Theta)\right) ,
\end{equation}
where $\omega_{3}$ is the Chern--Simons three-form for the relevant connection, for example
\begin{equation}
\omega_{3}(A) = \tr(A\wedge \dd A + \tfrac{2}{3}A\wedge A\wedge A) .
\end{equation}
This implies a non-trivial Bianchi identity for the NSNS three-form flux $H$
\begin{equation}
\dd H = \tfrac{1}{4}\alpha'(\tr F\wedge F - \tr R\wedge R) . 
\end{equation}
For convenience, in what follows we will drop explicit reference to $\alpha'$, absorbing it into the definition of $B$ and $\omega$. Moreover we will mostly ignore the tangent bundle connection $\Theta$ with the understanding that it can be reintroduced afterwards by taking $V$ to be a $G=\Gg\times \GL{6,\mathbb{R}}$ vector bundle, where $\Gg$ is the gauge group for $A$, together with a suitable definition of the trace, as, for example, in~\cite{Garcia-Fernandez:2013gja}.

It is useful to group these equations into so-called F-terms and D-terms. As was discussed in~\cite{OHS16}, the F-term conditions correspond to
\begin{equation}\label{eq:F-terms}
\dd(\ee^{-2\varphi}\Omega) = 0,\qquad
\ii(\partial-\bar{\partial})\omega = H,\qquad
F_{0,2}= 0.
\end{equation}
The remainder are the D-terms
\begin{equation}\label{eq:D-terms}
\dd(\ee^{-2\varphi}\omega\wedge \omega) =0,\qquad
\omega^{\sharp}\lrcorner F=0.
\end{equation}

One can view the F-terms as determining a holomorphic structure on a certain bundle $\mathcal{Q}$~\cite{OS14b}. The remaining D-term conditions -- a conformally balanced metric and polystability of $V$ -- must then imposed. More precisely one requires the bundle $\mathcal{Q}$ to be holomorphic, where $\mathcal{Q}$ is defined via a series of extensions as
\begin{equation}
\label{eq:holo-Courant}
\begin{array}{rcccl}
T^{*1,0}& \longrightarrow& \mathcal{Q}& \longrightarrow & \mathcal{A} , \\
\left(\ad P_{G}\right)_{\mathbb{C}} & \longrightarrow & \mathcal{A} & \longrightarrow & T^{1,0}.
\end{array}
\end{equation}
where $\ad P_{G}$ is a vector bundle with fibre $\mathfrak{g}$, the Lie algebra of the gauge group. This is an example of a holomorphic Courant algebroid~\cite{GRT18,GRST18}. Equivalently there exists a holomorphic differential $\bar{D}$ such that
\begin{equation}
\bar{D}\colon\Omega^{(p,q)}(X,\mathcal{Q}) \to \Omega^{(p,q+1)}(X,\mathcal{Q}) , \qquad \bar{D}^{2}=0.
\end{equation}
The condition $\bar{D}^{2}=0$ is equivalent to the integrability of the conventional complex structure, the holomorphicity of the gauge bundle and the Bianchi identities for $F$, $R$ and $H$\footnote{Note that this relies on the gauge group $G$ admitting a compact real form. More generally, this statement may not be true \cite{Garcia-Fernandez:2020awc}.}.

The moduli of the background appear in the massless spectrum of the four-dimensional theory and so a full knowledge of the moduli space is important for both phenomenology and more formal questions. Once one moves away from Calabi--Yau type solutions and allows non-zero fluxes, the moduli are much more difficult to understand. Fortunately, identifying the holomorphic structure $\bar{D}$ streamlines the analysis of the moduli space for heterotic compactifications~\cite{Anderson:2011ty,OS14b,OHS16,Ashmore:2018ybe}. The moduli can be thought of as deformations of $\bar{D}$ that still satisfy $\bar{D}^2=0$ and the D-term conditions. Given some mild assumptions on the bundle $V$, it is known~\cite{OS14b} that the hermitian Yang--Mills equations do not impose any extra conditions on the infinitesimal moduli of the system (and that the same result holds for $T$). It is also known that while deformations of the hermitian structure preserving the conformally balanced condition~\eqref{eq:conformally holomorphic-balanced} may a priori be infinite dimensional, once you impose the anomaly cancellation condition you are reduced to a finite number of moduli. Up to $(0,2)$ variations of the NSNS two-form $B$, the infinitesimal moduli of the Hull--Strominger system are then given by deformations of the holomorphic structure on $\mathcal{Q}$. That is they are counted by the cohomology
\begin{equation}
H^{0,1}_{\bar{D}}(X,\mathcal{Q}).
\end{equation}
We should note that these actually include non-physical moduli which correspond to deformations of the connection $\Theta$ that do not change the physical fields, such as the metric.\footnote{These are counted by $H_{\bar{\nabla}}^{(0,1)}(X,\End T)$, where $\bar{\nabla}$ is the antiholomorphic part of the covariant derivative defined by $\Theta$.} These appear in this construction as one treats $\Theta$ as an independent field (and part of the gauge connection), whereas in reality it is determined by the other fields of the background. To find the physical moduli, one must remove this over counting -- this has yet to be understood.

The story outlined above is valid for infinitesimal deformations. Using holomorphicity, one can also study finite deformations~\cite{Ashmore:2018ybe}. These are known to obey the Maurer--Cartan equation for an $L_{3}$ algebra (an $L_\infty$ algebra up to degree 3). The deformations can be packaged into
\begin{equation}
    y\in\Omega^{(0,1)}(X,\mathcal{Q}),\qquad b\in \Omega^{(0,2)}(X),
\end{equation}
where $y$ encodes deformations of the holomorphic structure -- deformations of the complex structure, complexified hermitian structure and gauge connection -- and $b$ encodes the $(0,2)$ deformations of the $B$ field. Note that the $b$ modulus vanishes if $h^{0,2}=0$~\cite{McOrist:2019mxh} -- we will make no such assumption here and so shall keep explicit reference to it. To linear order the moduli are determined by the set of equations
\begin{align}
\bar{D}y - \tfrac{1}{2}\partial b &= 0 \label{eq:AOMSCS 1}, \\
\bar{\partial} b &= 0 \label{eq:AOMSCS 2}, \\
\partial(\ee^{-2\varphi}\imath_{\mu}\Omega) &=0 \label{eq:AOMSCS 3},
\end{align}
where $\mu\in \Omega^{(0,1)}(X, T^{1,0})$ is a complex structure deformation. These are the equations we will recover in section~\ref{sec:moduli}.

\section{Generalised structures for \texorpdfstring{$\mathcal{N}=1$}{N=1} heterotic backgrounds}
\label{sec:Gen Geom for heterotic}

Generalised geometry provides a useful framework for studying generic supersymmetric backgrounds of maximal supergravities in terms of integrable generalised $G$-structures. In particular, it gives a geometric interpretation of generic properties of type II and M-theory backgrounds, such as the superpotential and Kähler potential for $\mathcal{N}=1$ solutions with four external dimensions, as well as tools to tackle questions about the moduli space~\cite{N1paper}. Heterotic (and type I) theories can also be formulated in terms of generalised geometry, as we will now summarise briefly. We will then discuss how generalised geometry can be used to characterise $\mathcal{N}=1$ heterotic backgrounds.

Ignoring the gauge bundle for now, the bosonic field content of the heterotic theory is the same as the NSNS sector of type II supergravity. Hence the relevant generalised geometry is that of $\Orth{6,6}\times\mathbb{R}^{+}$ generalised geometry on a generalised tangent bundle $E$ defined as an extension of $T$ by $T^{*}$~\cite{Hitchin02,Gualtieri04}
\begin{equation}
T^{*}\longrightarrow E' \longrightarrow T,
\end{equation}
where $E$ admits an $\Orth{6,6}\times\mathbb{R}^{+}$ structure. As usual, there is a natural differential operator known as the generalised Lie (or Dorfman) derivative on $E$. An (off-shell) configuration of the bosonic fields defines a generalised metric that reduces the structure group of $E$ to $\SO{6}\times\SO{6} \simeq \SU4\times \SU4$. 

We can reintroduce the gauge connection and obtain full heterotic backgrounds as follows. Combining the connection $\Theta$ with the gauge connection $A$ to give a single connection on the principal bundle $P_G$, where $G=\Gg\times\GL{6,\bbR}$, the generalised tangent bundle $E$ is defined as the extension
\begin{equation}
\begin{array}{ccccc}
T^{*} & \longrightarrow & E' & \longrightarrow & E ,\\
\ad P_{G} & \longrightarrow & E & \longrightarrow & T,
\end{array} \label{eq:holomorphic courant algebroid}
\end{equation}
where $\ad P_{G}$ is the vector bundle with fibre $\mathfrak{g}$, the Lie algebra of the extended gauge group $G$. This structure with its Dorfman derivative is known as a transitive Courant algebroid~\cite{severa-letter} -- it has been used to describe heterotic supergravity in~\cite{Garcia-Fernandez:2013gja,CMTW14} (see also \cite{Hohm:2011ex} in the double field theory context). We review some of the key points in appendix~\ref{ap:O(6,6+n) generalised geometry}. In particular, given a generalised vector $V\in\Gamma(E)$, there is a Dorfman derivative $\Dorf_V$ defined by \eqref{eq:het_dorfman}. Locally we have a (non-canonical) isomorphism
\begin{equation}
E\simeq T\oplus \ad P_{G} \oplus T^{*}. \label{eq:het gen tangent bundle}
\end{equation}
This has a natural $\Orth{6,6+n}$ structure on it defined by the inner product
\begin{equation}
\eta(v+\Lambda + \lambda,w + \Sigma +\sigma) = \tfrac{1}{2}\imath_{v}\sigma + \tfrac{1}{2}\imath_{w}\lambda +\tr(\Lambda\Sigma), \label{eq:O(6,6+n) structure}
\end{equation}
where $n$ is the dimension of $\mathfrak{g}$. While we will not give the exact form of the adjoint bundle $\ad\tilde{F}$ whose fibres are the Lie algebra $\orth{6,6+n}$, we note that
\begin{equation}
T^{*}\otimes \mathfrak{g} \subseteq \ad\tilde{F} \simeq \wedge^{2}E.
\end{equation}
An (off-shell) configuration of the bosonic fields, that is a metric $g$, two-form $B$ and one-form gauge field $A$, again define a generalised metric that in this case reduces the structure group to $\SO{6}\times\SO{6+n}$~\cite{Garcia-Fernandez:2013gja}. Further requiring the fields to give a solution preserving $\mathcal{N}=1$ supersymmetry is equivalent to a further reduction to an integrable $\SU3\times \SO{6+n}$ structure. As in previous work on $\mathcal{N}=1$ structures~~\cite{Ashmore:2018ybe}, we will find it useful to also consider a weaker $\mathbb{R}^{+}\times \Uni3\times \SO{6+n}$ structure. We will see how these are defined in terms of generalised structures in section \ref{sec:U(3)xSpin(6+n)} and how to define the conditions for integrability.

Note that in the formalism where one includes the $\Theta$ connection by extending the gauge bundle $V$ to be a $\Gg\times \GL{6,\mathbb{R}}$ bundle, there are non-physical degrees of freedom, since the connection $\Theta$ on the tangent space connection is thought of as independent of the metric and $B$. One can remove these by setting the value of $\Theta$ by hand. As was discussed in~\cite{CMTW14}, one can get around this issue by identifying an $\Orth6$ subbundle of the $\GL{6,\mathbb{R}}$ bundle, then identifying it with one of the $\Orth6$ structures defined on the $T\oplus T^{*}$ part of the generalised tangent bundle. This gives a structure group $\Orth6\times \Gg\times\Orth6$. The trade off is that the generalised connections relevant for this construction will not be torsion free, but instead appear with a particular non-vanishing intrinsic torsion. We will not take this approach in this paper.

\subsection{\texorpdfstring{$\SU3\times\SU4$}{SU(3)xSU(4)} and \texorpdfstring{$\bbR^+\times\Uni3\times\SU4$}{R+xU(3)xSU(4)} structures}\label{sec:U(3)xSU(4)}

Let us start by considering the simple case where we ignore the gauge bundle, applicable to both the heterotic and type II theories. As discussed in~\cite[Appendix C]{Ashmore:2018ybe}, the existence of a nowhere-vanishing spinor that can parametrise  $\mathcal{N}=1$ supersymmetry transformations in four dimensions requires a reduction of the structure group from that defined by the generalised metric, namely $\SU4\times\SU4$, to $\SU3\times\SU4\subset\Orth{6,6}\times\bbR^+$. Following~\cite{N1paper}, it will be useful for us to also define a slightly weaker $\mathbb{R}^{+}\times \Uni3 \times \SU4$ structure. These will play roles analogous to $\SL{3,\mathbb{C}}$ structures and $\GL{3,\mathbb{C}}$ structures in conventional geometry.

Each structure is defined by a generalised tensor that is invariant under the reduced structure group\footnote{Note that, as we will argue below, the particular determinant weight of the $\psi$ structure is required to make $\psi$ a holomorphic function on the space of $\SU3\times\SU4$ structures.}
\begin{equation}
\begin{aligned}
    \text{$\SU3\times\SU4$ structure :} && 
    \psi &\in\Gamma(\det T^* \otimes \ext^3E_\bbC ) , \\
    \text{$\UR$ structure :} && J &\in\Gamma(\ad\tilde{F}) . 
\end{aligned}
\end{equation}
They are stabilised by the same $\SU3\times\SU4$, but $J$ is also invariant under a $\mathbb{C}^*$ action. As discussed in detail in~\cite{N1paper}, one should think of this as generalising the relation between an $\SL{3,\mathbb{C}}$ structure $\Omega$ and a $\GL{3,\mathbb{C}}$ structure $I$. The differential conditions which ensure supersymmetry of the on-shell solution are then equivalent to the integrability of this structure, in line with the general discussion of~\cite{CSW14b}. In the next section we will see how we can reformulate the conditions for integrability of the $\mathbb{R}^{+}\times \Uni3\times\SU4$ structure, and in the following section consider the extra conditions that make the $\SU3 \times \SU4$ structure integrable. 

Let us begin by defining the structure $J$. At a point on the manifold, the generalised metric defines an $\SU4\times\SU4$ subgroup of $\Orth{6,6}\times\mathbb{R}^+$, with the invariant spinor reducing this further to $\SU3\times\SU4$. There is a $\Uni1\subset\SU4$ that commutes with the $\SU3$. The commutant of this $\Uni1$ inside $\Orth{6,6}\times\mathbb{R}^+$ is an $\mathbb{R}^+\times\Uni3\times\SU4$, where the $\Uni1$ is generated at each point of the internal manifold by a section $J\in\Gamma(\ad\tilde{F})$.\footnote{As in the type II and M-theory case~\cite{N1paper}, one can also define $J$ at each point on the manifold as being conjugate to a certain element of $\su4\times\su4$ that commutes with the desired $\su3\times\su4$.} This leads us to define
\begin{defn*}
A \emph{generalised $\,\mathbb{R}^+\times U(3)\times SU(4)$ structure} is a section $J\in\Gamma(\ad \tilde F)$ that generates this $\Uni1$ subgroup at each point.
\end{defn*}
\noindent
By construction, $J$ defines a generic reduction of the structure group of the generalised tangent bundle $E$ to $\mathbb{R}^+\times\Uni3\times\SU4$.\footnote{Note that the standard \emph{generalised complex structure}~\cite{Hitchin02,Gualtieri04} is also defined by choosing the generator of a $\Uni1$ subgroup but in that case the commutant would be $\Uni{3,3}$.} Different choices of $J$ are related by local $\Orth{6,6}\times\mathbb{R}^+$ transformations, giving an orbit of structures within the $\rep{66}$ representation space.

Decomposing $\Orth{6,6}$ using explicit $\SU4\times\SU4$ indices, we have
\begin{equation}
    \rep{66}=(\rep{15},\rep{1})\oplus(\rep{1},\rep{15})\oplus(\rep{6},\rep{6})\ni(\mu^\alpha{}_\beta,\mu^{\dot\alpha}{}_{\dot\beta},\mu^{\alpha\beta\dot\alpha\dot\beta}),
\end{equation}
where the nowhere-vanishing spinor $\epsilon$ is invariant under an $\SU3$ subgroup of the first $\SU4$ factor. Using this, we can write $J$ as
\begin{equation}
\label{eq:J-spinor-form}
    J^\alpha{}_\beta = 4\,\epsilon^\alpha \bar\epsilon_\beta-(\bar\epsilon \epsilon)
    \delta^\alpha{}_\beta, \qquad
    J^{\dot{\alpha}}{}_{\dot{\beta}} 
    = J^{\alpha\beta\dot{\alpha}\dot{\beta}} = 0 , 
\end{equation}
where we have normalised $\bar\epsilon \epsilon = 1$. Decomposing further under the $\SU3\times\Uni1$ subgroup of the first $\SU4$ factor, we have
\begin{equation}\label{eq:66}
    \rep{66}=(\rep{8},\rep{1})_0\oplus(\rep{3},\rep1)_{-2}\oplus(\repb3,\rep1)_{2}\oplus(\rep1,\rep1)_0\oplus(\rep1,\rep{15})_0\oplus(\rep3,\rep6)_{1}\oplus(\repb3,\rep6)_{-1},
\end{equation}
where a non-bold subscript denotes the $\Uni1$ charge. $J$ lies in the singlet $(\rep1,\rep1)_0$ representation. 

From the expression~\eqref{eq:J-spinor-form} and the parametrisation of the generalised metric in terms of a conventional metric $g$ and two-form field $B$, one finds that $J$ generically takes the form 
\begin{equation}
\label{eq:J-I-omega}
J=\tfrac{1}{2}\,\ee^{-B}\cdot(I-\omega+\omega^{\sharp}),
\end{equation}
where $I$ is the almost complex structure on $T_{\mathbb{C}}$ defined by the three-form $\Omega$, and $\omega$ is the compatible fundamental two-form. The $B$ field acts by the exponentiated adjoint action, which is nilpotent at degree three. In analogy with a conventional complex structure, we can use $J$ to decompose the generalised tangent space into eigenspaces. Under $\SU3\times\Uni1\times\SU4$, the adjoint action of $J$ on the complexification of $E$ splits as
\begin{equation}
\begin{split}
    E_\mathbb{C} &= L_1 \oplus L_{-1} \oplus L_0,\\
    \rep{12}_\mathbb{C} &= (\rep3,\rep1)_1\oplus(\repb3,\rep1)_{-1}\oplus(\rep1,\rep6)_0.\\
\end{split}
\end{equation}
Given the form~\eqref{eq:J-I-omega}, it is then easy to see that $L_{-1}$ takes the generic form 
\begin{equation}
L_{-1} = \ee^{-B-\ii\omega}\cdot T^{0,1} 
   = \{\bar{v}+\imath_{\bar{v}}(B+\ii\omega)\; | \; \bar{v}\in \Gamma(T^{0,1}) \} \label{eq:U(3)xSU(4) L_-1},
\end{equation} 
where as above $T^{0,1}\subset T_\bbC$ is the $-\ii$ eigenbundle for the action of the almost complex structure $I$.\footnote{We will denote $(0,1)$-vectors with a bar. Unbarred objects will denote either generic vectors or $(1,0)$-vectors depending on context. The complex conjugate of a vector or one-form will be indicated with a superscript ${}^*$.} As with a conventional almost complex structure, we have an alternative definition purely in terms of the subbundle $L_{-1}$:
\begin{defn*}
An $\mathbb{R}^+\times\Uni3\times\SU4$ structure is a subbundle $L_{-1}\subset E_\mathbb{C}$ such that
\begin{enumerate}
    \item[i)] $\dim_{\mathbb{C}}L_{-1}=3$,
	\item[ii)] $\eta(L_{-1},L_{-1})=0$,
	\item[iii)] $L_{-1}\cap\bar{L}_{-1}=\{0\}$,
	\item[iv)] The map $h\colon L_{-1}\times L_{-1}\rightarrow \bbC$, defined by $h(V,W)= \eta(V,\bar{W})$, is a definite hermitian inner product.
\end{enumerate}
\end{defn*}
\noindent
Note that we could equally well define the structure in terms of $L_1$.

Turning to the $\SU3\times\SU4$ structure $\psi$, we note that the bundle 
\begin{equation}
    K= \det T^*\otimes\ext^3E ,
\end{equation}
transforms in the $\rep{220}_{\rep1}$ representation of $\Orth{6,6}\times\mathbb{R}^+$ (where the bold subscript denotes the $\mathbb{R}^+$ weight~\cite{CSW11b}). Decomposing first under $\SU4\times\SU4$ and then under $\SU3\times\Uni1\times\SU4$, we have 
\begin{equation}
\begin{split}\label{eq:su3u1su4_decomposition}
    \rep{220}&=(\rep{10},\rep1)\oplus(\repb{10},\rep1)\oplus(\rep1,\rep{10})\oplus(\rep1,\repb{10})\oplus(\rep{15},\rep6)\oplus(\rep6,\rep{15})\\
    &= (\rep1,\rep1)_3\oplus(\repb6,\rep1)_{-1}\oplus(\rep3,\rep1)_1\oplus(\rep1,\rep1)_{-3}
    \oplus(\rep{6},\rep{1})_{1} \oplus (\bar{\rep{3}},\rep1)_{-1}\oplus\ldots.
\end{split}
\end{equation}
where the subscripts now denote the $\Uni1$ charge. In particular, we see that the $\SU3\times\SU4$ singlet in the decomposition implies that each choice of $J$ defines a unique line bundle $\mathcal{U}_J\subset K_\mathbb{C}$, satisfying
\begin{equation}\label{eq:L1_proj}
    V\bullet\psi = 0 \quad \forall\;V\in\Gamma(L_{-1}),\qquad \eta(\psi,\bar\psi)\neq 0,
\end{equation}
where $\psi$ is a local section of $\mathcal{U}_J$, $\eta$ is the pairing on sections of $K$ induced from the symmetric pairing $\eta$ on $E$, and the product $V\bullet\psi$ is the projection map $E\otimes K \to H$, where $H$ is the generalised tensor bundle transforming in the $\rep{495}_{\rep{1}}$ representation of $\Orth{6,6}\times\mathbb{R}^+$.
Equivalently, a local section $\psi$ is defined by $J\psi=-3\ii\psi$ under the adjoint action of $J$.\footnote{This corresponds to taking $\psi\in(\rep1,\rep1)_{-3}$. We make this choice to match with the usual conventions of $\Omega$ being the holomorphic object on the space of structures.} Mirroring the definition of a nowhere-vanishing three-form for an almost complex structure, we then have
\begin{defn*}
Given a choice of $J$ with trivial line bundle $\mathcal{U}_J$, a generalised $\SU3\times\SU4$ structure is a global nowhere-vanishing section $\psi\in\Gamma(\mathcal{U}_J)$.
\end{defn*}
\noindent
Note that two different choices of $\psi$ that are related by multiplication by a nowhere-vanishing complex function define the same structure $J$. Decomposing with explicit $\SU4\times\SU4$ indices we have 
\begin{equation}
\label{eq:220decomp}
\begin{split}
    \rep{220} &=\repp{10}{1} \oplus (\repb{10},\rep{1}) 
        \oplus \repp{15}{6} \oplus \repp{6}{15} 
        \oplus \repp{1}{10} \oplus (\rep{1},\repb{10}) \\
    & \ni(\kappa^{\alpha\beta},\kappa_{\alpha\beta},
        \kappa^\alpha{}_{\beta}{}^{\dot{\alpha}\dot{\beta}},
        \kappa^{\dot{\alpha}}{}_{\dot{\beta}}{}^{\alpha\beta},
        \kappa^{\dot{\alpha}\dot{\beta}},\kappa_{\dot{\alpha}\dot{\beta}}) .
\end{split}
\end{equation}
In terms of the spinor $\epsilon$ we then have 
\begin{equation}
\label{eq:psi-eta}
    \psi^{\alpha\beta} = \sqrt{g}\,\ee^{-2\varphi}\,\epsilon^\alpha \epsilon^\beta ,
\end{equation}
with all the other components vanishing. Recall that $\psi$ is defined up to a complex function. We fixed the normalisation $\bar{\epsilon}\epsilon=1$, so that the phase of $\epsilon$ encodes the phase freedom in $\psi$, while the overall scale of $\psi$ is parameterised by the dilaton $\ee^{-2\varphi}$, in line with the fact that the combination $\sqrt{g}\,\ee^{-2\varphi}$ is the $\Orth{6,6}$ invariant volume defined by the generalised metric~\cite{CSW11b}. 

Again we can use the generalised metric to translate this into a tensor expression following~\cite{Ashmore:2018ybe}. As we have mentioned a generalised metric gives a reduction of the structure group of $E$ to $\SO6_{+}\times\SO6_{-}\simeq\SU4_+\times\SU4_-$. The $\Orth{6,6}\times\bbR^+$ generalised tangent bundle $E$ then decomposes under $\SO6_{+}\times\SO6_{-}$ as $E = C_+ \oplus C_-$, giving a corresponding decomposition of $\ext^3 E$ as
\begin{equation}
\ext^3E = 
\ext^3C_+ \oplus (\ext^2C_+\otimes C_-) 
\oplus (C_+\otimes\ext^2C_-) \oplus \ext^3C_- ,
\end{equation}
as in~\eqref{eq:220decomp}, where the $\ext^3C_\pm$ spaces decompose into complex self-dual and anti-self-dual components transforming in the $\rep{10}$ and $\repb{10}$ representations. Note that, in terms of the splitting defined by the generalised metric we have
\begin{equation}
(C_{+})_{\mathbb{C}} = L_{1}\oplus L_{-1}, \qquad (C_{-})_{\mathbb{C}}=L_{0}.
\end{equation}
Let $\hat{E}^+_a=\hat{e}_a + e_a - \imath_{\hat{e}_a}B$ be an explicit basis for $C_+$, where $\hat{e}_a$ is an orthonormal basis for $T$ defined by the metric $g$, and $e_a$ is the dual basis. The expression~\eqref{eq:psi-eta} defines the tensor
\begin{equation}\label{eq:psi}
\begin{split}
\psi &= \sqrt{g}\,\ee^{-2\varphi}\,\tfrac{1}{3!} (\epsilon^\text{T}\gamma^{abc}\epsilon)
    \,\hat{E}^+_a \wedge \hat{E}^+_b \wedge \hat{E}^+_c \\
    &= \ee^{-2\varphi}\,\ee^{-B-\ii\omega}\cdot \Omega ,
\end{split}
\end{equation}
where the exponential $\ee^{-B-\ii\omega}$ acts via the adjoint action and in going to the second line we use the isomorphism $\ext^3 T \otimes \ext^6 T^*\simeq \ext^3 T^*$. This expression ensures $\psi$ is stabilised by the correct $\SU3\times\SU4$ subgroup. We note that given an $\mathcal{N}=2$ structure encoded by a pair of pure spinors $\Phi_\pm$, one can construct $\psi$ as
\begin{equation}
\psi^{MNP} = (\bar\Phi_+ , \Gamma^{MNP} \Phi_-),
\end{equation}
where $\Gamma^{M}$ are the $\Orth{6,6}$ gamma matrices and $(\cdot,\cdot)$ is the Mukai pairing.

\subsection{\texorpdfstring{$\SU3\times\Spin{6+n}$}{SU(3)xSpin(6+n)} and \texorpdfstring{$\mathbb{R}^{+}\times\Uni3\times\Spin{6+n}$}{R+xU(3)xSpin(6+n)} structures}\label{sec:U(3)xSpin(6+n)}

It is straightforward to extend this story to include the gauge bundle. Since many of the results are analogous to the previous section, we will sketch the key points. As noted in~\eqref{eq:het gen tangent bundle}, the generalised tangent bundle is locally given by
\begin{equation}
E \simeq T\oplus\ad P_G \oplus T^*,
\end{equation}
where $\ad P_G$ is the adjoint bundle with fibres given by the Lie algebra $\mathfrak{g}$ of the gauge group $G$. Sections of $E$ thus encode diffeomorphisms and gauge transformations of both the gauge field $A$ and the two-form $B$. Again, there are two generalised structures each defined by a generalised tensor that is invariant under the reduced structure group
\begin{equation}
\begin{aligned}
    \text{$\SU3\times\Spin{6+n}$ structure :} && 
    \psi &\in\Gamma(\det T^* \otimes \ext^3E_\bbC ) , \\
    \text{$\mathbb{R}^{+}\times \Uni3\times\Spin{6+n}$ structure :} && J &\in\Gamma(\ad\tilde{F}) . 
\end{aligned}
\end{equation}
These are stabilised by the same $\SU3\times\Spin{6+n}$, but $J$ is also invariant under a $\mathbb{C}^*$ action.

We begin with the weaker $\mathbb{R}^{+}\times\Uni3\times\Spin{6+n}$ structure defined by $J$. Mirroring the discussion in the previous subsection, one finds that $J$ generically takes the form
\begin{equation}
\label{eq:J-I-omega-gauge}
J=\tfrac{1}{2}\,\ee^{-B}\ee^{-A}\cdot(I-\omega+\omega^{\sharp}),
\end{equation}
where now we include a twisting by the one-form gauge field $A$. Again, we can use $J$ to decompose the generalised tangent space $E$ into eigenspaces. Noting that the fibres of $E$ transform in the $(\rep{12}+\rep{n})$ representation of $\Orth{6,6+n}$ and decomposing under $\Uni1\times\SU3\times \Spin{6+n}$ we find that
\begin{equation}
\begin{split}
    E_\mathbb{C} &= L_1 \oplus L_{-1} \oplus L_0,\\
    \rep{12}+\rep{n} &= (\rep3,\rep1)_{1}+(\repb3,\rep1)_{-1}+(\rep1,\rep{6}+\rep{n})_{0},\\
\end{split}
\end{equation}
where $(\rep6+\rep n)$ is the fundamental representation of $\Spin{6+n}$. Identifying $L_{-1}$ as the subbundle transforming as $(\repb3,\rep1)_{-1}$, given the form of $J$ in \eqref{eq:J-I-omega-gauge}, one can check that $L_{-1}$ takes the generic form
\begin{equation}
L_{-1}=\ee^{-B-\ii\,\omega}\ee^{-A}T^{0,1}=\{\bar{v}+\imath_{\bar{v}} A	+\imath_{\bar{v}}(B+\ii\,\omega)- \tr(\imath_{\bar{v}} A \,A)\; | \; \bar{v}\in \Gamma(T^{0,1})\}, \label{eq:U(3)xSpin(6+n) L_-1}
\end{equation}
where $T^{0,1}\subset T_\bbC$ is the $-\ii$ eigenbundle for the almost complex structure $I$. As before, one can use $L_{-1}$, subject to some algebraic conditions, as a definition of the $\mathbb{R}^{+}\times\Uni3\times\Spin{6+n}$ structure.

As in the case without the gauge bundle, an $\SU3\times\Spin{6+n}$ structure $\psi$ is a nowhere-vanishing section of
\begin{equation}
\psi \in \Gamma(\det T^* \otimes\ext^{3}E_\mathbb{C}).
\end{equation}
Again, $\psi$ is not a generic element but needs to lie in a particular orbit of $\Spin{6,6+n}$ so that its stabiliser is $\SU3\times\Spin{6+n}$. Using a generalised metric, we can write $E=C_{+}\oplus C_{-}$, where $C_{+}$ is a six-dimensional subbundle on which $\eta$ is positive definite, defined in \eqref{eq:O(6,6+n) structure}. Letting $\hat{E}^{+}_{m}$ be a basis for $C_{+}$, we can write
\begin{equation}
\begin{split}\label{eq:SU(3)xSpin(6+n) psi}
\psi &= \sqrt{g}\ee^{-2\varphi} \tfrac{1}{3!}(\epsilon^\text{T}\gamma^{mnp}\epsilon)\hat{E}^{+}_{m}\wedge\hat{E}^{+}_{n}\wedge\hat{E}^{+}_{p}\\
&= \ee^{-2\varphi}\ee^{-B-\ii\omega}\ee^{-A}\cdot\Omega . 
\end{split}
\end{equation}
This expression guarantees that $\psi$ is stabilised by the correct $\SU3\times\Spin{6+n}$ group.

\subsection{Supersymmetry and intrinsic torsion}\label{sec:susy_int}

The existence of the $\psi$ structure is just the algebraic part of the supersymmetry conditions for an $\mathcal{N}=1$ background (namely the requirement that one has a non-vanishing spinor). There are also differential conditions given by the Killing spinor equations, which can be translated into the F- and D-term conditions in~\eqref{eq:F-terms} and~\eqref{eq:D-terms} respectively. As we will discuss, these are equivalent to the structure being torsion-free or ``integrable''~\cite{Garcia-Fernandez:2013gja,CMTW14,CSW14b}. As in~\cite{N1paper}, it will be useful to consider the intrinsic torsion for both $J$ and $\psi$ as one can view an integrable $\psi$ in terms of an integrable $J$ together with a further differential condition in the form of a moment map for generalised diffeomorphisms.

We call a structure torsion-free or integrable if there exists a generalised connection that is compatible with the structure and is torsion-free. For example, a torsion-free $\SU3\times\Spin{6+n}$ structure is equivalent to the existence of $\psi$ and a connection $\Dgen$ such that
\begin{equation}\label{eq:torsion_def}
\Dgen\psi = 0, \qquad L^{\Dgen}_{V} - L_{V} = T(V)=0,
\end{equation}
where $L_{V}$ is the Dorfman derivative defined in~\eqref{eq:het_dorfman} with the gauge sector turned off, $L^{\Dgen}_{V}$ is the Dorfman derivative with $\partial$ replaced by $\Dgen$, and the generalised torsion is a map $T\colon \Gamma(E)\to\Gamma(\ad\tilde{F})$. The obstruction to the existence of such a torsion-free connection is a non-vanishing intrinsic torsion. 

Starting with the simpler case where we ignore the gauge bundle, following the standard analysis~\cite{CSW11b,CSW11,CSW14b,AW15,CS16}, we find that the intrinsic torsion for $\psi$ and $J$ live in subbundles of $\ext^3 E \oplus E^*$ transforming as $\rep{220}\oplus\rep{12}$ and decomposing under the structure group via
\begin{align}
\begin{split}\label{eq:su3_su4_int_tor}
W^{\text{int}}_{\SU3\times\SU4} &: (\rep3,\rep6)_{-2} \oplus (\repb3,\rep6)_{2}\oplus(\rep1,\rep1)_{-3} \oplus (\rep1,\rep1)_{3} \\
& \eqspace \oplus (\rep3,\rep1)_{1} \oplus (\repb3,\rep1)_{-1}\oplus(\rep1,\rep6)_{0},
\end{split}\\
W^{\text{int}}_{\mathbb{R}^{+}\times\Uni3\times\SU4} &: (\rep3,\rep6)_{-2} \oplus (\repb3,\rep6)_{2}\oplus(\rep1,\rep1)_{-3} \oplus (\rep1,\rep1)_{3},\label{eq:r_u3_su4_int_tor}
\end{align}
where the subscript denotes the $\Uni1$ charge under $J$. 

When we include the gauge bundle the representations in which the intrinsic torsion for each structure lives are given by
\begin{align}
\begin{split}\label{eq:SU(3)xSpin(6+n) int torsion}
W^{\text{int}}_{\SU3\times\Spin{6+n}} &: (\rep3,\rep6+\rep n)_{-2} \oplus (\repb3,\rep6+\rep n)_{2} \oplus (\rep1,\rep1)_{-3} \oplus (\rep1,\rep1)_{3} \\
& \eqspace \oplus (\rep3,\rep1)_{1} \oplus (\repb3,\rep1)_{-1} \oplus (\rep1,\rep6+\rep n)_{0},
\end{split}\\
W^{\text{int}}_{\mathbb{R}^{+}\times\Uni3\times\Spin{6+n}} &: (\rep3,\rep6+\rep n)_{-2} \oplus (\repb3,\rep6+\rep n)_{2} \oplus (\rep1,\rep1)_{-3} \oplus (\rep1,\rep1)_{3} ,\label{eq:U(3)xSpin(6+n) int torsion}
\end{align}
where a subscript denotes the $\Uni1$ charge with respect to $J$ and $(\rep6+\rep n)$ is the fundamental representation of $\Spin{6+n}$.

Since $\mathcal{N}=1$ supersymmetry in four dimensions follows from integrability of the $\SU3\times\SU4$ structure, and integrability is equivalent to the vanishing of the intrinsic torsion of the structure, we need some natural differential conditions which enforce the vanishing of the above components of the intrinsic torsion. These differential conditions should then be thought of as the supersymmetry conditions for the background, but now with a geometric interpretation. The form of these conditions will be the subject of the next two sections. 

\section{Involutivity, the superpotential and F-terms}
\label{sec:involutivity}


In this section we will consider the integrability of the weaker $\mathbb{R}^{+}\times\Uni3\times\SU4$ and $\bbR^+\times\Uni3\times\Spin{6+n}$ structures, defined by $J$, and the show how these conditions can be defined as an involutivity condition of a subbundle or equally as coming from varying a superpotential. This matches an earlier observation, in the case of pure $\Orth{d,d}$ generalised geometry, relating
supersymmetry of the underlying sigma model to integrability of a subbundle~\cite{Hull:2018jkr}. We will also briefly discuss the connection to the holomorphic Courant algebroid~\cite{OS14b,GRST18,GRT18} given in~\eqref{eq:holo-Courant}. We will turn to the extra conditions that one must impose on $\psi$ to guarantee an honest $\mathcal{N}=1$ background in the next section. 

\subsection{Involutivity conditions}

As with conventional complex structures and the $\mathcal{N}=1$ structures defined in \cite{N1paper}, it turns out that integrability of the $J$ structure is equivalent to involutivity of a subbundle of the generalised tangent bundle. For the $\UR$ structure we define
\begin{defn*}
A torsion-free $\mathbb{R}^{+}\times\Uni3\times\SU4$ structure $J$ is one for which $L_{-1}$ is involutive under the Dorfman derivative
\begin{equation}
    \Dorf_V W \in \Gamma(L_{-1})\qquad \forall\;V,W\in\Gamma(L_{-1}).
\end{equation}
\end{defn*}
\noindent
Note that one can replace the Dorfman derivative with the Courant bracket in this condition: the difference between the two is a term of the form $\dd(\eta(V,W))$, but $\eta(V,W)$ vanishes for $V,W\in\Gamma(L_{-1})$ from the definition of an $\mathbb{R}^+\times\Uni3\times\SU4$ structure. We also note that since $\bar{L}_{-1}\simeq L_1$, involutivity of $L_{-1}$ is equivalent to involutivity of $L_1$.

It is straightforward to see that involutivity of $L_{-1}$ is equivalent to vanishing intrinsic torsion for the $\mathbb{R}^{+}\times\Uni3\times\SU4$ structure. Recall first that we can always find a generalised connection $D$ that is compatible with the structure, so that $DJ=0$, but this is not necessarily torsion-free. Now consider the definition \eqref{eq:torsion_def} of the torsion of a connection where we restrict to $V,W\in\Gamma(L_{-1})$
\begin{equation}
L_{V}W = L_{V}^{\Dgen}W - T(V)\cdot W \label{eq:U(3)xSU(4) integrability 1}.
\end{equation}
Compatibility of the connection guarantees $\Dorf_V^\Dgen W\in\Gamma(L_{-1})$, so involutivity reduces to checking that $T(V)\cdot W$ lies only in $L_{-1}$. Note also that since the left-hand side does not depend on the choice of connection and $L_{V}^{\Dgen}W$ lies in $\Gamma(L_{-1})$ for any choice of $\Dgen$, only the intrinsic torsion can contribute to the components of $T(V)\cdot W$ that lie outside of $L_{-1}$. The intrinsic torsion representations that appear in $T(V)\cdot W \in \Gamma(E)$ are
\begin{equation}
    \begin{split}
        (\repb3,\rep6)_{2}\otimes(\repb3,\rep1)_{-1} \otimes (\repb3,\rep1)_{-1} &\supset (\rep1,\rep6)_0,\\
        (\rep1,\rep1)_{3}\otimes(\repb3,\rep1)_{-1} \otimes (\repb3,\rep1)_{-1} &\supset (\rep3,\rep1)_{1}.
    \end{split}
\end{equation}
A non-zero $(\repb3,\rep6)_{2}$ component of the intrinsic torsion would generate a $(\rep1,\rep6)_0\simeq L_0$ term in $\Dorf_V W$, while a non-zero $(\rep1,\rep1)_{3}$ component would generate a $(\rep3,\rep1)_{1}\simeq L_1$ part. Requiring both of these to be absent so that $L_V W \in \Gamma(L_{-1})$ only sets both of these components of the intrinsic torsion to zero. Complex conjugation then implies that the whole of the intrinsic torsion vanishes. This shows that the $\mathbb{R}^{+}\times\Uni3\times\SU4$ structure defined by $J$, or equivalently $L_{-1}$, is integrable if and only $L_{-1}$ is involutive with respect to the Dorfman derivative.

The discussion up to this point has been rather abstract. One might wonder how integrability for $J$ translates into concrete equations for the $\SU3$ structure that underlies the Hull--Strominger system discussed in section \ref{sec:Hull-Strominger Review}. Given that we have an explicit description of the subbundle $L_{-1}$, given in \eqref{eq:U(3)xSU(4) L_-1}, we can check how involutivity constrains the $\SU3$ structure. Taking any $ v, w\in\Gamma(T)$ one finds
\begin{equation}
\Dorf_{\ee^{-B-\ii\,\omega}v}(\ee^{-B-\ii\,\omega} w)=\ee^{-B-\ii\,\omega}\Dorf^{H+\ii\,\dd\omega}_{v} w=\ee^{-B-\ii\,\omega}\bigr([v,w]-\imath_{v}\imath_{w}(H+\ii\,\dd\omega)\bigr),
\end{equation}
where $H=\dd B$ and we have used the expression for the Dorfman derivative in~\eqref{eq:het_dorfman} after setting the gauge field to zero. If in particular we choose the vectors to be $\bar{v},\bar{w}\in\Gamma(T^{0,1})$ so that $\ee^{-B-\ii\,\omega}\bar{v}\in\Gamma(L_{-1})$, then for $L_{-1}$ to be involutive (so that the right-hand side lies only in $L_{-1}$), we require that $[\bar v,\bar w]-\imath_{\bar v}\imath_{\bar w}(H+\ii\,\dd\omega)$ is a section of $\Gamma(T^{0,1})$ alone. Splitting into vector and one-form equations, this gives the conditions
\begin{equation}
[\bar v,\bar w]  \in\Gamma(T^{0,1}),\qquad
\imath_{\bar v}\imath_{\bar w}(H+\ii\,\dd\omega)=0,
\end{equation}
which must hold for all choices of $\bar v,\bar w\in\Gamma(T^{0,1})$. The first of these is simply the requirement that the almost complex structure $I$ is integrable, so that it is an honest complex structure. This also implies that the corresponding complex three-form $\Omega$ satisfies $\dd\Omega=\bar{a}\wedge\Omega$ for some $\bar{a} \in \Omega^{0,1}(X)$. The second condition can be understood by decomposing according to complex type as $H=H_{3,0}+H_{2,1}+H_{1,2}+H_{0,3}$ and $\omega=\omega_{1,1}$. Since $\bar v$ and $\bar w$ are $(0,1)$-vectors, the second of the conditions gives $H_{0,3}=0$ and $H_{1,2}+\ii\,\bar{\partial}\omega=0$. As both $H$ and $\omega$ are real, these imply $H_{3,0}=H_{0,3}=0$ and $H_{2,1}+H_{1,2}+\ii(\bar{\partial}-\partial)\omega=0$. Putting this together, we have
\begin{equation}
L_{-1}\text{ is involutive}\quad\Leftrightarrow\quad\begin{aligned} [\bar v,\bar w]  & \in\Gamma(T^{0,1})\\
H&=\ii(\partial-\bar{\partial})\omega
\end{aligned}
\end{equation}
Note that these are (almost) the equations coming from the F-term conditions~\eqref{eq:F-terms} with the gauge bundle turned off. The F-term equations are slightly stronger since they imply that $\Omega$ is conformally holomorphic, fixing $\bar a$ in terms of the dilaton $\varphi$, whereas the above conditions leave $\bar a$ undetermined. We will come back to this point when we discuss the superpotential in section~\ref{sec:superpotential}. Note also that these are the same set of conditions as the integrability of a ``half generalised complex structure''~\cite{Hull:2019iof}, which appear from a worldsheet analysis of $(2,0)$ non-linear sigma model geometry.

The involutivity condition naturally extends to the $\bbR^+\times\Uni3\times\Spin{6+n}$ case. Given the explicit description of $L_{-1}$ in \eqref{eq:U(3)xSpin(6+n) L_-1} and the expression for the Dorfman derivative in \eqref{eq:het_dorfman}, we can relate integrability for the $\mathbb{R}^{+}\times\Uni3\times\Spin{6+n}$ structure, in the form of involutivity of $L_{-1}$, to the data of the Hull--Strominger system, namely the $\SU3$ structure and the connection on $V$. Taking generic vectors $v,w\in\Gamma(T)$ one now finds
\begin{equation}\label{eq:het_dorf}
\Dorf_{\ee^{-B-\ii\,\omega}\ee^{-A}v}(\ee^{-B-\ii\,\omega}\ee^{-A}w)=\ee^{-B-\ii\,\omega}\ee^{-A}\bigl([v, w]-\imath_{v}\imath_{w}(H+\ii\,\dd\omega)-\imath_{v}\imath_{w}F\bigr),
\end{equation}
where
\begin{alignat}{8}
H  &=\dd B+\omega_{3}(A), \qquad&
\omega_{3}(A)  &=\tr(A\wedge\dd A+\tfrac{2}{3}A\wedge A\wedge A),\\
F &=\dd A+A\wedge A,& \dd H &= \tr(F\wedge F).
\end{alignat}
As before, specialising to $\bar{v},\bar{w}\in\Gamma(T^{0,1})$ so that $\ee^{-B-\ii\,\omega}\ee^{-A}\bar{v} \in \Gamma(L_{-1})$, for involutivity of $L_{-1}$ we require that the expression in the parentheses in~\eqref{eq:het_dorf} lies only in $\Gamma(T^{0,1})$. This implies
\begin{equation}
L_{-1}\text{ is involutive}\quad\Leftrightarrow\quad\begin{aligned} [\bar v,\bar w] & \in\Gamma(T^{0,1})\\
H & =\ii(\partial-\bar{\partial})\omega\\
F_{0,2} & =0
\end{aligned}
\end{equation}
As before, we have an integrable complex structure on the manifold, implying $\dd\Omega=\bar{a}\wedge\Omega$ for some $\bar{a}\in\Omega^{0,1}(X)$, and the three-form flux $H$ is fixed by $\dd^I$ of the hermitian form $\omega$. In addition, the $(0,2)$ component of the curvature $F$ must vanish, implying that the gauge bundle is holomorphic. Again, these are the F-term equations~\eqref{eq:F-terms}, up to the conformal holomorphicity condition for $\Omega$.

In order to describe the heterotic theory, as mentioned, we can include the tangent bundle connection within the gauge sector, as discussed in \cite{Garcia-Fernandez:2013gja,CMTW14,GRST18}. This has the effect of redefining $H$ to be its full heterotic form and adds a holomorphicity condition for the tangent bundle connection so that
\begin{equation}
H  =\dd B+\omega_{3}(A)-\omega_{3}(\Theta),\qquad
R_{0,2}  =0,
\end{equation}
where $\Theta$ is the $\nabla^{-}$ connection and $R$ is the corresponding curvature two-form.

It is interesting to compare how the involutivity condition on $L_{-1}$ defines the holomorphic structure of the geometry to the holomorphic Courant algebroid $\mathcal{Q}$ given in~\eqref{eq:holo-Courant} and used in the papers~\cite{GRT18,GRST18}. Defining the perpendicular subbundle $L_{-1}^\perp$, such that, on a patch $U_i$,
\begin{equation}
    V \in \Gamma(L_{-1}^\perp) \qquad \Leftrightarrow \qquad \eta(V,W)=0 \quad \forall\; W\in\Gamma(L_{-1}) , 
\end{equation}
we have 
\begin{equation}
\begin{split} \label{eq:holomorphic algebroid from L}
    L_{-1}^\perp \quotient L_{-1} 
      &\simeq \ee^{-B-\ii\omega}\ee^{-A}\cdot\left( 
          T^{1,0} \oplus T^{0,1} \oplus T^{*1,0} \oplus (\ad P_{G})_\mathbb{C}\right) \quotient
          \ee^{-B-\ii\omega}\ee^{-A}\cdot T^{0,1} , \\
      &\simeq \ee^{-B-\ii\omega}\ee^{-A}\cdot\left( 
          T^{1,0} \oplus T^{*1,0} \oplus (\ad P_G)_{\mathbb{C}}\right) , \\
      &\simeq  T^{1,0} \oplus T^{*1,0} \oplus (\ad P_G)_{\mathbb{C}} 
      \simeq \mathcal{Q} . 
\end{split}
\end{equation}
Hence we see that $L_{-1}$ indeed determines $\mathcal{Q}$ and furthermore the involutivity of $L_{-1}$ implies that $\mathcal{Q}$ is holomorphic.\footnote{Note that it is the adjoint bundle for the complexified group, $G_\mathbb{C}$, that appears here. If $L_{-1}$ is involutive, so that we have $F_{0,2}=0$, the transition functions that define $(\ad P_G)_{\mathbb{C}}$ can be taken to be holomorphic, so that $\mathcal{Q}$ is also holomorphic.} As a bundle, all $\mathcal{Q}$ are isomorphic to  $T^{1,0} \oplus T^{*1,0} \oplus (\ad P_G)_\mathbb{C}$. However the corresponding holomorphic Courant algebroids (or more precisely ``Bott--Chern algebroids'' in the language of~\cite{GRST18}) are distinguished by the choice of $\omega$ and $A$, such that inequivalent algebroids are distinguished by the Aeppli class defined in~\cite{GRST18}.

\subsection{The superpotential}\label{sec:superpotential}

It is known that the F-term conditions in \eqref{eq:F-terms} can be derived starting from a heterotic superpotential~\cite{GLM04,Benmachiche:2008ma,OHS16,McOrist16}
\begin{equation}\label{eq:het_superpotential}
\mathcal{W}=\int_X \ee^{-2\varphi}\Omega\wedge(H+\ii\,\dd\omega),
\end{equation}
and requiring $\mathcal{W}=\delta\mathcal{W}=0$ under variations of the structures $\Omega$, $\omega$ and fields $B$ and $\varphi$~\cite{OHS16,McOrist16}. Building on work on flux superpotentials~\cite{GVW00,Gukov:1999gr} and their description in generalised geometry~\cite{PW08}, we conjectured in~\cite{N1paper} that the superpotential is given by the singlet part of the intrinsic torsion of the $\psi$ structure and explicitly showed this was true for the examples of $\Gx2$ in M-theory and generic $\mathcal{N}=1$ backgrounds of type II theories. Here we will show that the singlet torsion does indeed give the superpotential in the case of heterotic backgrounds and that it is a holomorphic function of $\psi$. We also discuss how the superpotential conditions imply involutivity of $L_{-1}$. Not only does this provide a covariant expression for the superpotential for generic heterotic backgrounds, it also provides further justification for the claim made in~\cite{N1paper}.

Given that an infinitesimal change in $\psi$ can be parametrised by an element of the $\Orth{6,6+n}\times\bbR^+$ Lie algebra and $\psi$ transforms in the $\repp{1}{1}_{-3}$, the variations of the $\SU3\times\Spin{6+n}$ structure $\psi$ transform as $(\rep1,\rep1)_{-3}$, $(\repb3,\rep1)_{-1}$ and $(\rep3,\rep{6+n})_{-2}$. Thus $\delta\mathcal{W}/\delta\psi=0$ constrains the dual $(\rep1,\rep1)_{3}$, $(\rep3,\rep1)_{1}$ and $(\repb3,\rep{6+n})_{2}$ components of the intrinsic torsion. Note that this means the vanishing of the variation of $\mathcal{W}$ implies $\mathcal{W}=0$, as $\mathcal{W}$ is the singlet component of the intrinsic torsion. We also note that the superpotential condition is slightly stronger than involutivity of $L_{-1}$, which constrained only the $(\rep1,\rep1)_{3}$ and $(\repb3,\rep{6+n})_{2}$ components, leaving $(\rep3,\rep1)_{1}$ undetermined. The involutivity condition implies there is an integrable complex structure and hence $\dd\Omega=\bar a \wedge \Omega$. The extra superpotential constraint is precisely what is needed to fix the $(0,1)$-form $\bar a$. 

As for $\Ex{7(7)}\times\mathbb{R}^+$ backgrounds with $\mathcal{N}=1$ supersymmetry, one can rephrase involutivity as a holomorphic condition on $\psi$ itself. Let $V\in\Gamma(L_{-1})$ and $D$ be a compatible connection, such that $D\psi=0$. From the definition of the torsion of $D$ in \eqref{eq:torsion_def}, we have
\begin{equation}
    \Dorf_V \psi = -T(V)\cdot\psi\qquad\text{for }V\in\Gamma(L_{-1}).
\end{equation}
Naively one would expect $\Dorf_V^D\psi$ to appear on the right-hand side. This would contain terms of the form $D_V\psi$, $(D\times_{\ad}V)\cdot\psi$ and $(D\cdot V)\psi$ (where the final term appears as $\psi$ has a non-zero weight under the $\mathbb{R}^+$ action). However, using the fact that $\psi$ is a singlet and that it has weight one under $\mathbb{R}^+$, one finds that the terms which involve $D$ acting on $V$ cancel identically, leaving only $D_V \psi$ which vanishes due to the compatibility of the connection. The remaining torsion term is linear in $V$ and, since $\Dorf_V \psi$ is independent of $D$, only the intrinsic torsion can appear in $T(V)\cdot\psi$. Using the $\Uni1\times\SU3\times\SU4$ decomposition, one can check that the $(\repb3,\rep{6+n})_{2}$, $(\rep1,\rep1)_{3}$ and $(\rep3,\rep1)_{1}$ parts of the intrinsic torsion \eqref{eq:su3_su4_int_tor} appear, which are the same components that appear in $\delta\mathcal{W}/\delta\psi$. This gives us an alternative description of the involutivity condition as
\begin{equation}\label{eq:inv_hol}
    \text{involutive }L_{-1} \qquad \Leftrightarrow\qquad \Dorf_V \psi = U(V)\,\psi\quad\forall\;V\in \Gamma(L_{-1}),
\end{equation}
where $U\in\Gamma(L_{-1}^*)$ is the $(\rep3,\rep1)_{1}$ component of the $\SU3\times\SU4$ intrinsic torsion, and $U(V)=U_M V^M$ is a pairing between sections of $E^*$ and $E$ so that $U(V)$ is a scalar function. If we further require that $U$ vanishes, we have
\begin{equation}
    \frac{\delta\mathcal{W}}{\delta\psi} = 0 \qquad \Leftrightarrow \qquad \Dorf_V \psi =0 \quad \forall\;V\in \Gamma(L_{-1}),
\end{equation}
so that we have an alternative description of the superpotential condition (recall that $\delta \mathcal{W}/\delta\psi=0$ implies $\mathcal{W}=0$). As discussed in \cite{N1paper}, we expect that one can take a given $\psi$ that satisfies the involutivity condition and rescale it by an appropriate complex function so that the stronger superpotential condition is satisfied. Note that these expressions show that involutivity and the superpotential itself are holomorphic in $\psi$. Since $L_{-1}$ is fixed by $V\bullet\psi=0$ (see \eqref{eq:L1_proj}), $L_{-1}$ depends holomorphically on $\psi$. The conditions that $\Dorf_V \psi=U(V)\psi$ and $\Dorf_V \psi=0$ for all $V\in\Gamma(L_{-1})$ are then also holomorphic in $\psi$ (since $\bar\psi$ does not appear).


Our conjecture that the superpotential is given by the singlet of the intrinsic torsion can be translated to the statement that
\begin{equation}
    \mathcal{W}=\int_X W \sim \int_X \eta(\psi,T),
\end{equation}
where $T$ is the intrinsic torsion of the structure. The pairing of $T$ with $\psi$ projects onto the $(\rep1,\rep1)_{3}$ component. Note also that $\psi$ is weight one and $T$ is weight zero under the $\mathbb{R}^+$ action, so that their pairing is a weight-one scalar. A weight-one scalar is a section of $\det T^*$ and so gives a volume form that can be integrated over the manifold. From the previous discussion, the $(\rep1,\rep1)_{3}$ component of the torsion can be obtained from $\psi$ alone, and so the superpotential itself is a holomorphic function of $\psi$.

There are alternative ways to write $\mathcal{W}$ to make the dependence on $\psi$ more obvious. One can always find a torsion-free connection $\Dgen$ that is compatible with the generalised metric structure discussed in section \ref{sec:U(3)xSpin(6+n)}. Using this one can write the integrand of the superpotential as
\begin{equation}
\label{eq:superpotential density2}
W \sim \tr(J,\Dgen\times_{\ad}\psi),
\end{equation}
where $J$ is the $\mathbb{R}^{+}\times \Uni3\times\Spin{6+n}$ structure defined in section \ref{sec:U(3)xSpin(6+n)}.\footnote{As for the case of $\Ex{7(7)}\times\bbR^+$ generalised geometry~\cite{N1paper}, it is easy to see that this expression does not depend on the choice of connection (such torsion-free compatible connections are not unique). In particular, there are no singlets in the undetermined parts of $\Dgen$ when one decomposes under the $\cN=1$ structure group. This means that any expression that is an $\SU3 \times \Orth{6+n}$ singlet, is linear in $\Dgen$ and involves only $\SU3 \times \Orth{6+n}$ invariant tensors, will depend only on the singlet part of the $\SU3 \times \Orth{6+n}$ intrinsic torsion.} Note that since neither $J$ nor the generalised connection are weighted under $\mathbb{R}^{+}$, the right-hand side of \eqref{eq:superpotential density2} is a section of $\det T^*$ and hence we can integrate it over the manifold to give
\begin{equation}
\mathcal{W} 
\sim\int_{X} \tr(J,\Dgen\times_{\ad}\psi)\label{eq:superpotential}.
\end{equation}
This expression is the easiest to use for direct calculations. Naively it does not appear to be holomorphic in $\psi$ as $J$ is a function of $\psi$ and $\bar\psi$. However, we can rewrite it as
\begin{equation}
\mathcal{W} \sim\int_{X}  \frac{\eta( \bar{\psi},(\Dgen\times_{\ad}\psi)\cdot\psi )}{\eta( \bar{\psi},\psi )}, \label{eq:superpotential2}
\end{equation}
where, as in \cite{N1paper}, the weight of $\psi$ is such that the dependence on $\bar\psi$ drops out. That is, under an infinitesimal antiholomorphic variation of $\bar{\psi}$, only the terms that are proportional to $\bar{\psi}$ contribute to the variation of $\eta( \bar{\psi},(\Dgen\times_{\ad}\psi)\cdot\psi )$, while the other components are projected out. This leaves a trivial scaling transformation $\bar{\psi}\rightarrow \ee^{\bar{c}}\bar{\psi}$, under which our expression is clearly invariant thanks to $\eta( \bar{\psi},\psi )$ in the denominator. Hence $\mathcal{W}$ does not vary under deformations of $\bar{\psi}$ and so it is indeed holomorphic in $\psi$, as we claimed.

As we show in appendix \ref{sec:explicit_moment}, using the explicit expressions for $J$ and $\psi$ in terms of the underlying $\SU3$ structure, the superpotential reduces to
\begin{equation}
\mathcal{W} \sim \int_{X}\ee^{-2\varphi}\Omega\wedge(H+\ii\,\dd\omega).
\end{equation}
This is precisely the form of the superpotential in \eqref{eq:het_superpotential} and used in \cite{Benmachiche:2008ma,OHS16,McOrist16}. Hence our expression \eqref{eq:superpotential} is the covariant form of the superpotential for a generic four-dimensional $\mathcal{N}=1$ heterotic background determined by $\psi$.

Having seen how the F-term conditions of the Hull--Strominger system can be understood as involutivity for a subbundle defined by a generalised structure or the vanishing of the superpotential, in the next section we will discuss how the remaining D-term equations can be imposed by requiring the vanishing of a moment map for generalised diffeomorphisms. This moment map will be defined using $\psi$, and its vanishing will be equivalent to the vanishing of the remaining components of the intrinsic torsion for the $\SU3\times\Spin{6+n}$ structure, confirming the claim that a four-dimensional $\mathcal{N}=1$ heterotic background is equivalent to an integrable $\SU3\times\Spin{6+n}$ structure.

\section{The Kähler potential, moment map and D-terms}\label{sec:Kahler, moment and D}

As we have seen, integrability of the $\Uni 3\times\Spin{6+n}\times\mathbb{R}^{+}$ structure – in the form of involutivity of $L_{-1}$ – gives a subset of the supersymmetry conditions required of an $\mathcal{N}=1$, $D=4$ heterotic background. As we have mentioned, the remaining conditions come from the vanishing of a moment map for the action of diffeomorphisms and gauge transformations (generalised diffeomorphisms). Much of what follows is analogous to the story for $\Ex{7(7)}\times\mathbb{R}^+$ backgrounds. For this reason, we shall be brief and refer the interested reader to the longer discussion in \cite{N1paper}.

\subsection{The Kähler potential}

We know that the moduli space of a generic four-dimensional $\mathcal{N}=1$ theory admits a Kähler metric which will be related to the Kähler poential on the space of $\SU3\times\Spin{6+n}$ structures. Here we will give an expression for this potential in terms of the object $\psi$.

At each point $p\in X$, $\psi$ is stabilised by some $\SU3\times\Spin{6+n}\subset \Orth{6,6+n}\times\mathbb{R}^{+}$ subgroup. Hence at each point, $\psi$ is an element of the coset
\begin{equation}
\psi|_{p}\in \Q= \frac{\Orth{6,6+n} \times\mathbb{R}^{+}} {\SU3\times\Spin{6+n}}.
\end{equation}
An $\SU3\times\Spin{6+n}$ structure is then a section of the fibre bundle
\begin{equation}
\Q \longrightarrow \QQ\longrightarrow X.
\end{equation}
Hence we can define the space of $\SU3\times\Spin{6+n}$ structures to be the set of sections of $\QQ$:
\begin{equation}
\mathcal{Z}\simeq\Gamma(\QQ).
\end{equation}
There is a natural Kähler structure on this space, determined by supersymmetry. First, note that the homogeneous space $\Orth{6,6+n}/\Uni3\times\Spin{6+n}$ admits a pseudo-Kähler structure~\cite{Borel54}. The space $\QQ$ can be viewed as a complex line bundle over this homogeneous space with the zero section removed. This reflects the fact that we only have an $\mathbb{R}^{+}$ action, and hence we have a cone over a Kähler base. This complex cone over a Kähler base has a natural Kähler structure which then induces one on the space of sections. In this case, the Kähler potential $\mathcal{K}$ on $\mathcal{Z}$ is given by
\begin{equation}
\mathcal{K}=\int_{X}\eta(\psi,\bar{\psi})^{\frac{1}{2}}\label{eq:kahler potential},
\end{equation}
where $\psi$ is viewed as a complex coordinate on the space of structures. Note that the weight of $\psi$ ensures that $\eta(\psi,\bar{\psi})^{1/2}$ is a top-form and hence can be integrated. Different choices of weight would correspond to different Kähler metrics, with the weight we have chosen corresponding to the metric picked out by supersymmetry (as we saw with holomorphy of the superpotential).

As was shown in \cite{Ashmore:2018ybe}, the object $\psi$ does indeed give a complex coordinate on $\mathcal{Z}$. The particular form of $\psi$ and its $\mathbb{R}^+$ weight turns out to be very natural. Consider the anchor map 
\begin{equation}
\pi \colon E \to T,
\end{equation}
which simply projects on the vector component of a generalised vector. This induces a map $\pi \colon \ext^3 E \to \ext^3 T$ which, together with $\ext^3 T \otimes \ext^6 T^*\simeq \ext^3 T^*$, gives
\begin{equation}
\pi(\psi) \sim \ee^{-2\varphi}\, \Omega.
\end{equation}
Thus, via the anchor map, the object $\psi$ defines an ordinary complex three-form $\pi(\psi)$ on the manifold. This three-form is $\Omega$ up to a dilaton factor, and is precisely the form that is holomorphic (closed under $\bar\partial$) in the Hull--Strominger system \eqref{eq:conformally holomorphic-balanced}. Note that, so long as we consider only deformations fixing the cohomology of the $H$ flux, we are fixing the underlying Courant algebroid and thus the anchor map $\pi$. The induced map is therefore complex linear and has no moduli dependence. This means that if $\psi$ is holomorphic on the coset $\Q$ then so is the three-form $\ee^{-2\varphi}\, \Omega$.

We can define a \emph{non-holomorphic} coordinate on $\mathcal{Z}$ as
\begin{equation}
\chi = \eta(\psi,\bar{\psi})^{-1/4}\psi .
\end{equation}
This is a complex section of $\ext^{3}E\otimes(\det T^*)^{1/2} \sim \rep{220}_{\rep{1}/\rep{2}}$ and gives the Kähler potential \eqref{eq:kahler potential} as
\begin{equation}
\mathcal{K} = \int_{X}\eta(\chi,\bar{\chi}).
\end{equation}
We will see that this non-holomorphic parametrisation is useful for writing the symplectic structure on $\mathcal{Z}$. The symplectic structure on $\mathcal{Z}$ is given by $\varpi=\ii\,\partial'\bar\partial'\mathcal{K}$, where $\delta=\partial'+\bar\partial'$ is the functional derivative on $\mathcal{Z}$. Contracting two vectors $\alpha,\beta\in\Gamma(T\mathcal{Z})$ into $\varpi$, one has
\begin{equation}
\begin{split}
    \imath_\beta\imath_\alpha\varpi &= \frac{\ii}{2}\int_X \eta(\psi,\bar\psi)^{-1/2}\Bigl(\eta(\imath_\alpha\delta\psi,\imath_\beta\delta\bar\psi)-\eta(\imath_\beta\delta\psi,\imath_\alpha\delta\bar\psi)\\
    &\eqspace\phantom{\frac{\ii}{2}\int_X}-\tfrac{1}{2}\eta(\psi,\bar\psi)^{-1}\eta(\imath_\alpha\delta\psi,\bar\psi)\eta(\psi,\imath_\beta\delta\bar\psi)+\tfrac{1}{2}\eta(\psi,\bar\psi)^{-1}\eta(\imath_\beta\delta\psi,\bar\psi)\eta(\psi,\imath_\alpha\delta\bar\psi)\Bigr).
\end{split}
\end{equation}
Rewriting this in terms of $\chi$ gives
\begin{equation}
    \imath_\beta \imath_\alpha \varpi = \frac{\ii}{2} \int_X \left(\eta(\imath_\alpha\delta \chi,\imath_\beta \delta\bar\chi)-\eta(\imath_\beta \delta\chi,\imath_\alpha \delta\bar\chi)\right).
\end{equation}

While we leave the full calculation to appendix \ref{sec:explicit_moment}, one can show that the Kähler potential takes the form
\begin{equation}\label{eq:potential_omega}
\mathcal{K} = \int_X\ii\,\ee^{-2\varphi}\Omega\wedge\bar{\Omega}.
\end{equation}
In fact, it takes this form up to an overall constant which can be removed by rescaling $\psi$. With this rescaling $\chi$ is given by
\begin{equation}\label{eq:chi_def}
\chi = \frac{1}{3!}g^{1/4}\ee^{-\varphi}\Omega^{mnp} \hat{E}^{+}_{mnp},
\end{equation}
where $\hat{E}^{+}_{mnp}=\hat{E}^{+}_{m}\wedge \hat{E}^{+}_{n}\wedge \hat{E}^{+}_{p}$, and the $\hat{E}^{+}_{m}$ are defined as in \eqref{eq:SU(3)xSpin(6+n) psi}. We will see later that, while \eqref{eq:potential_omega} appears to only depend on the complex structure parameters (which vary $\Omega$), it does in fact capture all possible deformations of the structure.

\subsection{The moment map}\label{sec:moment}

One can then restrict to the subspace of $\psi$ structures for which $L_{-1}$ is involutive, that is
\begin{equation}
    \hat{\mathcal{Z}}=\{\psi\in\mathcal{Z}\;|\;J\text{ is integrable}\}.
\end{equation}
As we showed in \eqref{eq:inv_hol} in the discussion of the superpotential, this condition is holomorphic in $\psi$. Hence $\hat{\mathcal{Z}}$ inherits its Kähler metric $\mathcal{Z}$, which is defined by the same Kähler potential. Following the discussion in \cite{N1paper}, one can then define a moment map for the action of generalised diffeomorphisms on $\hat{\mathcal Z}$ as follows. Infinitesimally, generalised diffeomorphisms are generated by the Dorfman derivative along a generalised vector $V\in\Gamma(E)$. A generalised diffeomorphism defines a deformation of $\chi$ as
\begin{equation}
    \imath_{\rho_V}\delta\chi=\Dorf_V \chi,
\end{equation}
where $\rho_V\in\Gamma(T\hat{\mathcal Z})$ is the induced vector field. The corresponding moment map is defined by
\begin{equation}
    \imath_{\rho_V}\imath_\alpha \varpi = \imath_\alpha \delta\mu(V),
\end{equation}
from which we deduce
\begin{equation}\label{eq:moment_map}
    \mu(V) = - \frac{\ii}{2}\int_{X} \eta(\psi,\bar{\psi})^{-1/2} \eta(\Dorf_V \psi,\bar{\psi}) =  -\frac{\ii}{2} \int_X \eta(\Dorf_V \chi,\bar{\chi}),
\end{equation}
where $\mu\colon\hat{\mathcal Z} \to \mathfrak{gdiff}^*$ is the moment map. We will use the form of the moment map in terms of both $\psi$ and $\chi$ in the following, so we give them both above.

How does the moment map constrain the structure? In other words, which components of the intrinsic torsion can appear in $\mu$? Recall that we can always find a compatible connection ($\Dgen\psi=\Dgen\chi=0$) that is not necessarily torsion free. Using this we can rewrite the moment map as
\begin{equation}
\mu(V) =-\frac{\ii}{2}  \int_{X}\eta(L_{V}^{\Dgen}\chi,\bar{\chi}) +\frac{\ii}{2}  \int_{X}\eta(T^{\text{int}}(V)\cdot \chi,\bar{\chi}). \label{eq:moment map}
\end{equation}
The first term vanishes by the compatibility of $\Dgen$. Assuming that the associated weaker $\mathbb{R}^{+}\times\Uni3\times\Spin{6+n}$ structure is integrable, and hence its intrinsic torsion \eqref{eq:U(3)xSpin(6+n) int torsion} vanishes, the final term is zero for all $V\in \Gamma(E)$ if and only if the $(\rep3,\rep1)_{1}+(\repb3,\rep1)_{-1}+(\rep1,\rep6+\rep n)_{0}$ part of the intrinsic torsion in \eqref{eq:SU(3)xSpin(6+n) int torsion} vanishes.\footnote{Checking that $\mu(V)=0$ for all $V$ is equivalent to showing $\mu$ itself vanishes.} That is, imposing that the moment map vanishes, $\mu=0$, gives the final condition for the $\SU3\times\Spin{6+n}$ structure to be integrable. We then have
\begin{defn*}
A torsion-free generalised $\SU3\times\Spin{6+n}$ structure is one where the associated subbundle $L_{-1}$ is involutive and the moment map \eqref{eq:moment_map} vanishes.
\end{defn*}

We now check that the vanishing of the moment map imposes the remaining equations of the Hull--Strominger system that do not appear in the involutivity conditions found in the previous section. Taking a generic generalised vector $V=\ee^{-B}\ee^{-A}(v+\lambda+\Lambda)$ where $v\in\Gamma(T)$, $\lambda\in \Gamma(T^{*})$ and $\Lambda\in \Gamma(\ad P_{G})$, a long calculation in appendix~\ref{sec:explicit_moment} shows that
\begin{equation}
\begin{split}
\mu(V) &= \tfrac{1}{2}\int_X \imath_v(2\partial\varphi - 2\bar{\partial}\varphi + \bar{a} - a)\ee^{-2\varphi}\Omega\wedge\bar{\Omega} - 4 \,\ee^{-2\varphi}\tr(\Lambda F)\wedge \omega \wedge \omega \\
& \phantom{-\frac{\ii}{2}\int_X} +2 \lambda\wedge\dd (\ee^{-2\varphi}\omega\wedge \omega),
\end{split} \label{eq:D_terms_from_moment_map}
\end{equation}
where we have used the fact that the complex structure is integrable (which comes from involutivity) and so $\dd\Omega = \bar{a}\wedge\Omega$ for some $\bar a \in \Omega^{0,1}(X)$. It is clear that imposing the vanishing of the moment map for all $V=\ee^{-B}\ee^{-A}(v+\lambda+\Lambda)$ gives
\begin{equation}
\bar{a} = 2\bar{\partial}\varphi, \qquad F\wedge\omega \wedge \omega = 0, \qquad \dd(\ee^{-2\varphi}\omega\wedge \omega) = 0,
\end{equation}
which are equivalent to
\begin{equation}
\dd(\ee^{-2\varphi}\Omega) = 0, \qquad \omega^{\sharp}\lrcorner F = 0, \qquad \dd(\ee^{-2\varphi}\omega \wedge \omega) = 0.
\end{equation}
These are precisely the missing supersymmetry equations. Hence the Hull--Strominger system is equivalent to an integrable $\SU3\times\Spin{6+n}$ structure.

Physically, $\SU3\times\Spin{6+n}$ structures that are related by diffeomorphisms and gauge transformations (GDiff) give equivalent backgrounds, so the moduli space of structures $\mathcal{M}_\psi$ should be viewed as the space of torsion-free $\SU3\times\Spin{6+n}$ structures quotiented by the action of these transformations. Since $\hat{\mathcal Z}$ admits both a symplectic structure and a Kähler structure, there are two ways to view this quotient, namely as a symplectic quotient by $\GDiff$ or as a standard quotient by the complexified group $\GDiff_\mathbb{C}$:
\begin{equation}\label{eq:psi_moduli_space}
\mathcal{M}_\psi = \{\psi\in\hat{\mathcal Z}\;|\; \mu=0\}\quotient\GDiff\equiv\hat{\mathcal Z}\qquotient\GDiff \simeq \hat{\mathcal Z}\quotient\GDiff_\mathbb{C}.
\end{equation}
How is $\mathcal{M}_\psi$ related to the moduli space of $D=4$, $\mathcal{N}=1$ heterotic backgrounds? First note that even within $\mathcal{M}_\psi$, different choices of $\psi$ can lead to the same background, that is, the same set of physical fields.\footnote{Without the gauge sector, this is the statement that there is a family of $\psi$'s that give the same $\Orth{6}\times\Orth{6}$ structure.} Instead, it is the generalised metric that determines the physical fields, so we should take the moduli space of the background to be choices of $\psi\in\mathcal{M}_\psi$ that lead to different generalised metrics. Said differently, while deformations of $\psi$ at a point take values in $\Orth{6,6+n}\times\mathbb{R}^+/(\SU3\times\Orth{6+n})$, only those that are also in $\Orth{6,6+n}\times\mathbb{R}^+/(\Orth{6}\times\Orth{6+n}$ change the physical fields. Fortunately, it is easy to take this into account. First note that constant shifts of the dilaton can be absorbed in the definition of the four-dimensional metric (recall that we are working in string frame). Second, note that a deformation of $\psi$ that lives in $(\Spin{6}\times\Spin{6+n}/(\SU3\times\Spin{6+n})$ would correspond to a change of the Killing spinor $\epsilon$ that leaves the physical background unchanged. Such deformations are possible only if there is a second Killing spinor to rotate into, and so the background would secretly preserve $\mathcal{N}=2$ supersymmetry. Notice however that changes of $\epsilon$ by a constant phase do not lead to extra Killing spinors and such a phase can be absorbed into the four-dimensional spinors appearing in the split of the ten-dimensional spinor. This constant phase corresponds to the $\Uni1$ generated by $J$. Putting this together, assuming we do have an honest $\mathcal{N}=1$ background, the unphysical deformations of $\psi$ come from constant shifts of the dilaton and constant phase rotations. Given the form of $\psi$ in \eqref{eq:SU(3)xSpin(6+n) psi}, a constant shift of the dilaton by $\varphi\to\varphi - c/2$ simply rescales by the exponentiated $\mathbb{R}^+$ action of $c$ on a weight-one object. The physical moduli space $\mathcal{M}$ of the background is then
\begin{equation}
    \text{Moduli space of $\mathcal{N}=1$ background, }\mathcal{M}=\mathcal{M}_\psi \qquotient \Uni1 \simeq \mathcal{M}_\psi\quotient\mathbb{C}^*, \label{eq:physical moduli space}
\end{equation}
where $\lambda\in\mathbb{C}^*$ acts as $\psi\to\lambda\psi$. Note that this implies the Kähler potential scales as $\mathcal{K}\to|\lambda|\mathcal{K}$. The Kähler potential $\tilde{\mathcal{K}}$ on the physical moduli space is then
\begin{equation}\label{eq:physical_kahler}
    \tilde{\mathcal K} = - 3 \log \mathcal{K}.
\end{equation}

We can compare this expression with those found in the literature. The generic form of the Kähler potential, given an arbitrary (conventional) $\SU3$ structure, in the heterotic theory was given in~\cite{Benmachiche:2008ma} following~\cite{GLMW02,GLM04,GLM07} and for generic heterotic vacua in~\cite{COM17,McOrist16} (matching the original expressions in the case of Calabi--Yau compactifications~\cite{Strominger85,Dixon:1989fj,CL91}).  One finds 
\begin{equation}
   \tilde{\mathcal{K}} =-\log \mathcal{V}-\log(S+\bar S) - \log\int_X\ii \,\Psi\wedge\bar\Psi,
\end{equation}
where $\mathcal{V}$ is the volume calculated from $\omega$, $\re S\propto \ee^{-2\varphi}\mathcal{V}$ and $\Psi\propto\ee^{-2\varphi}\Omega$. Using the $\SU3$ structure relations and that the dilaton is independent of the internal manifold, one can rewrite the above expression as
\begin{equation}
\begin{split}
\tilde{\mathcal{K}}&=-\log \mathcal{V}-\log(\ee^{-2\varphi}\mathcal{V}) - \log \ee^{-4\varphi}\mathcal{V}\\
&=-\log(\ee^{-6\varphi}\mathcal{V}^3)\\
&=-3\log\int_X \ii \,\ee^{-2\varphi}\Omega\wedge\bar\Omega.
\end{split}
\end{equation}
This matches both the form of $\mathcal{K}$ that we give above and confirms the coefficient of $-3$ in moving from the Kähler potential $\mathcal{K}$ on the moduli space of $\SU3\times\Spin{6+n}$ structure to the Kähler potential $\tilde{\mathcal K}$ on the physical moduli space, as mentioned around \eqref{eq:physical_kahler}. 

When one has an honest Calabi--Yau background, the Kähler potential can be separated into terms that give the metric for complex structure, Kähler and bundle moduli, plus a universal term for the dilaton. On a general $\mathcal{N}=1$ background, such a split is not possible and one simply has~\eqref{eq:kahler potential}. This also explains another possible point of confusion. Looking at~\eqref{eq:potential_omega}, one might be tempted to think that it depends only on complex structure parameters (which vary $\Omega$). However, this is an artifact of expressing the general form of the Kähler potential~\eqref{eq:kahler potential} at a chosen point on the parameter space. Variations of the Kähler potential should be written in terms of variations of the full structure $\psi$, and not simply $\Omega$, and then one will capture all of the possible deformations. Put another way, in writing~\eqref{eq:kahler potential} we have not picked out the holomorphic parameterisation of $\psi$.\footnote{Note that even in the Calabi--Yau case, the Kähler potential is naively independent of the gauge field moduli. However, the holomorphic Kähler moduli are shifted relative to the naive ones, and once these are picked out the dependence on the gauge moduli becomes explicit~\cite{Derendinger:1985cv}.}

\subsection{Extremisation of the Kähler potential and GIT}\label{sec:GIT}

As we have seen, the Hull--Strominger system is equivalent to the existence of an involutive subbundle and the vanishing of a moment map for generalised diffeomorphisms. However, as for the $\ExR{7(7)}$ backgrounds discussed in~\cite{N1paper}, the vanishing of the moment map is equivalent to extremising the Kähler potential over complexified generalised diffeomorphisms simply because $\hat{\mathcal{Z}}$ is Kähler~\cite{Wess:1992cp}. This reformulation allows us to make a direct connection to the work of~\cite{GRST18}. 

If we take $\mathcal{I}$ to be the complex structure on $\hat{\mathcal{Z}}$, then the action of complexified generalised diffeomorphisms are generated by $\rho_{V}\in \Gamma(T\hat{\mathcal{Z}})$ and $\mathcal{I}\rho_{W}\in \Gamma(T\hat{\mathcal{Z}})$. Since $\psi$ is a holomorphic coordinate on the space of structures, we have
\begin{equation}
    \mathcal{L}_{\mathcal{I}\rho_{V}}\psi = \imath_{\mathcal{I}\rho_{V}}\partial'\psi = \ii\,\imath_{\rho_{V}}\partial'\psi = \ii\, L_{V}\psi,
\end{equation}
where $\mathcal{L}$ is the Lie derivative on $\hat{\mathcal{Z}}$, and we have split the exterior (functional) derivative into holomorphic and antiholomorphic parts $\delta = \partial'+\bar{\partial}'$. Varying the Kähler potential along the orbit of an imaginary GDiff, we have
\begin{align}
    \begin{split}
    \mathcal{L}_{\mathcal{I}\rho_{V}}\mathcal{K} &= \frac{1}{2}\int_{X} \eta(\psi, \bar{\psi})^{-1/2}\bigl[ \eta(\imath_{\mathcal{I}\rho_{V}}\delta\psi,\bar\psi) + \eta(\psi,\imath_{\mathcal{I}\rho_{V}}\delta\bar{\psi})\bigr] \\
    &= \frac{\ii}{2}\int_{X} \eta(\psi, \bar{\psi})^{-1/2}\bigl[ \eta(\Dorf_V \psi,\bar\psi) - \eta(\psi,\Dorf_{V}\bar{\psi})\bigr] \\
    &= \ii \int_{X} \eta(\psi,\bar{\psi})^{-1/2}\eta(\Dorf_{V}\psi,\bar{\psi}) \\
    &= -2\,\mu(V).
    \end{split}
\end{align}
Thus we can think of the D-terms as coming from the vanishing of a moment map, or, since $\mathcal{K}$ is invariant under the real group $\GDiff$, the extremisation of the Kähler potential with respect to $\GDiff_\mathbb{C}$. 

In the work of~\cite{GRST18}, the Hull--Strominger system is viewed as extremising a ``dilaton functional'' over variations of the holomorphic Courant algebroid~\eqref{eq:holo-Courant} with fixed Aeppli class. We note first that the dilaton functional is precisely the Kähler potential defined above. Moreover, as discussed around~\eqref{eq:holomorphic algebroid from L}, the involutive bundle $L_{-1}$ defines the holomorphic Courant algebroid $\mathcal{Q}$ with a hermitian metric $(\omega,A)$\footnote{This is labelled $(\omega,\theta^{h})$ in the language of \cite{GRST18}.} defining a given Aeppli class. The authors of~\cite{GRST18} show that the variations within a fixed Aeppli class are given by\footnote{This is given by $\delta\omega = \ii \,c(h^{-1}\delta h, F_{h})+\partial\xi^{0,1}+ \overline{\partial\xi^{0,1}}$ in the language of \cite{GRST18}.} 
\begin{equation}
    \delta \omega = 2\tr(\theta F) + \partial \xi^* + \bar{\partial}\xi,  \qquad \delta A_{0,1} = -\bar{\partial}_{A}\theta.
\end{equation}
Examining equations \eqref{eq:het_exact_1}--\eqref{eq:het_exact_4}, one sees that these are precisely the transformations generated by $\ee^{-B-\ii\omega}\ee^{-A}(-\ii\xi+\ii\xi^*+\theta)\in\gdiff_{\mathbb{C}}$. Hence, extremising the dilaton functional follows directly from our picture of extremising the Kähler potential over complex generalised diffeomorphisms. Interestingly, we have a larger set of variations which are not included in those considered in~\cite{GRST18}, namely variations parameterised by some complex vector field $v\in \Gamma(T_{\mathbb{C}})\simeq \mathfrak{diff}_\mathbb{C}$. As shown in \eqref{eq:D_terms_from_moment_map}, it is these variations that ensure $\ee^{-2\varphi}\Omega$ is a holomorphic section (closed under $\bar\partial$). As shown in~\cite{GRST18}, provided such a section exists the variational problem of the dilaton functional is equivalent to the Hull--Strominger system. In our formulation however, the existence of a holomorphic volume form becomes part of the variational problem and does not need to be implemented by hand.

The present work also answers a question posed in \cite{GRST18}, namely whether there exists a moment map interpretation of the Hull--Strominger system. Furthermore, this interpretation provides a fascinating link with geometric invariant theory (GIT).\footnote{See \cite{Thomas06} and references therein for a review of GIT.} As in many other classic problems (including the hermitian Yang--Mills equations~\cite{Atiyah57,Donaldson85,UY86,UY89} and the equations of Kähler--Einstein geometry \cite{Fujiki90,Donaldson97,KEproof}), we can view the space of integrable $\SU3\times \Spin{6+n}$ structures as a quotient by a complexified group of some infinite-dimensional space of structures. Geometric invariant theory then tells us that we should identify
\begin{equation} \label{eq:polystable_moduli}
    \hat{\mathcal{Z}}\qquotient \GDiff \simeq \hat{\mathcal{Z}}^{\text{ps}}\quotient \GDiff_{\mathbb{C}},
\end{equation}
where $\hat{\mathcal{Z}}^{\text{ps}}$ is the subspace of $\hat{\mathcal{Z}}$ of ``polystable points''.  This arises as it is not guaranteed that all $\GDiff_{\mathbb{C}}$ orbits will intersect with the surface $\mu^{-1}(0)$. If an orbit does not intersect this surface, we call the points along it unstable and these are not included in $\hat{\mathcal{Z}}^{\text{ps}}$. By understanding which points are polystable, one would be able to relate the existence of solutions to a differential equation, namely $\mu=0$, to the algebraic data of the complex orbits. In~\eqref{eq:psi_moduli_space} we skipped over this subtlety of having to restrict to a subspace of $\hat{\mathcal{Z}}$ as it turns out that it is not be relevant for the infinitesimal moduli problem in section~\ref{sec:moduli}.

The standard procedure for identifying which points in $\hat{\mathcal{Z}}$ are polystable runs as follows. One considers $\Uni1\subset\GDiff$ actions generated by some $\rho_{V}\in \Gamma(T\hat{\mathcal{Z}})$. Under complexification we get some $\mathbb{C}^{*}\subset \GDiff_{\mathbb{C}}$ action, $\psi\rightarrow\psi(\nu), \, \nu\in \mathbb{C}^{*}$, and we consider the limit $\nu\rightarrow 0$. If there is a limiting point in $\hat{\mathcal{Z}}/\bbC^*$ (for example if the latter space was compact, which however is not that case here)
then in the limit the $\mathbb{C}^{*}$ action should coincide with the rescaling action
\begin{equation}
    \lim_{\nu\rightarrow0}\psi(\nu)=\nu^{w(\psi,V)}\psi_{0}
\end{equation}
for some $\psi_{0}\in \hat{\mathcal{Z}}$. Here $w(\psi,V)\in \mathbb{Z}$ is called the weight, and is quantised because we have a $\Uni1$ action. In this limit we also find that
\begin{equation}
    \lim_{\nu\rightarrow0}\mathcal{K}(\nu) = |\nu|^{w(\psi,V)}\mathcal{K}_{0}.
\end{equation}
By considering all possible $\Uni1\subset \GDiff$ subgroups, or one-parameter-subgroups, one then defines
\begin{equation}
    \begin{aligned}
        & \text{if $w(\psi,V)<0$ for all 1-PS then $\psi$ is stable,} \\
        & \text{if $w(\psi,V)\leq0$ for all 1-PS then $\psi$ is semistable,} \\
        & \text{if $w(\psi,V)>0$ for some 1-PS then $\psi$ is unstable.} 
    \end{aligned}
\end{equation}

The usual argument for the correspondence~\eqref{eq:polystable_moduli} relies on the ``norm functional'' (in this case the Kähler potential) being convex over the action of $\GDiff_{\mathbb{C}}$. This then ensures that there is a unique minimum of the functional, i.e.~a point where $\mu=0$, within the complex orbit of the stable points. However, as is pointed out in~\cite{GRST18}, there are concave orbits given by primitive deformations of $\omega$. Therefore, there may be multiple points along a given $\GDiff_{\mathbb{C}}$ orbit for which $\mu=0$ and so the correspondence \eqref{eq:polystable_moduli} may be more subtle. Despite this, understanding polystability should give us conditions for the existence of solutions to the Hull--Strominger system, if not uniqueness.


It is interesting to consider this constraint for $\Uni1$ subgroups of the gauge group $G$, generated by some $\theta\in\Gamma(\ad P_{G})$.\footnote{Note that here $\theta$ is an honest gauge parameter and not a section of the generalised tangent space.} First note that we can express the weight as follows
\begin{equation}
    w(\psi,V)\mathcal{K}_{0} = \mathcal{L}_{\mathcal{I}\rho_{V}}\mathcal{K}_{0} = - 2\,\mu_{0}(V),
\end{equation}
where $\mu_0(V)$ is the moment map evaluated on $\psi_0$. Hence we can define $\psi$ to be semistable if $\mu_{0}(V)\geq 0$. 
In order to lift the generator of the $\Uni1$ action into a generalised vector we take, as usual, $V=\ee^{-B}\ee^{-A}\theta=\ee^{-A}\theta$, then from~\eqref{eq:D_terms_from_moment_map}, we have 
\begin{equation}
    \mu(\theta)\sim \int_{X}\ee^{-2\varphi}\tr(\theta F)\wedge\omega\wedge\omega .  \label{eq:HYM weight}
\end{equation}
For $\varphi=0$, this is precisely the expression for the weight for the GIT problem associated to the hermitian Yang--Mills equations. The requirement that (in an appropriate limit)~\eqref{eq:HYM weight} is greater than or equal to zero for all possible $\theta$ has been shown to be equivalent to the slope stability of the gauge bundle $P\rightarrow M$. (See, for example, \cite{trautwein2015survey} for a review.) More generally, for conformally balanced hermitian metrics, in our case when $\dd(\ee^{-2\varphi}\omega\wedge\omega)=0$, a theorem of Buchdahl and Li--Yau~\cite{Buchdahl1988,YL87} states that solutions of the hermitian Yang--Mills equations require slope stability with respect to $\ee^{-2\varphi}\omega\wedge\omega$, precisely the combination that appears in our weight expression. Note that here the balanced condition actually comes from extremizing the Kähler potential under the action of complex one-form gauge transformations of $B$, so it would be a consequence of our more general stability condition.  


This, of course, requires further investigation.
For the moment, we content ourselves with pointing out that gauge conditions resembling slope stability appear naturally in the GIT picture, and that by understanding the constraints coming from all possible $\Uni1$ subgroups, one might be able to characterise polystability for the full Hull--Strominger system. Note for example, we could consider circle actions on the manifold generated by some vector field $\xi\in \Gamma(T)$. One might expect those coming from Hamiltonian symplectomorphisms to be related to the picture of Calabi--Yau stability developed in~\cite{Fujiki90,Donaldson97}.

\section{Moduli}\label{sec:moduli}

We will now analyse the massless moduli of a generic heterotic background in terms of some cohomological structure. We have seen that the conditions for a $D=4,\, \mathcal{N}=1$ Minkowski background can be rephrased in terms of integrable $\SU3\times \Spin{6+n}$ structures. By using this language we will be able to give a new interpretation to previous results found on infinitesimal moduli~\cite{OS14b,Ashmore:2018ybe}. We will follow the methods of \cite{N1paper} closely.

As discussed around \eqref{eq:physical moduli space}, the physical moduli space is given by
\begin{equation}
    \mathcal{M} = \mathcal{M}_{\psi}\quotient \mathbb{C}^{*} \qquad \mathcal{M}_{\psi} = \{\psi\,|\, J \text{ is integrable}\}\qquotient \GDiff \simeq \hat{\mathcal{Z}}\quotient \GDiff_{\mathbb{C}}.
\end{equation}
Writing the moduli space in this way greatly simplifies the deformation theory. First, relating the symplectic quotient to a complex quotient means that we do not need to solve the moment map condition. Instead, we need only consider deformations of $\psi$ that preserve the involutivity of $L_{-1}$, up the action of complexified generalised diffeomorphisms. Second, those elements of $\GDiff_\bbC$ that preserve $J$ simply rescale $\psi$ by a function. The moment map fixes this factor, up to an overall constant $\bbC^*$ rescaling. Thus we can actually identify the moduli space simply as a quotient of the space of integrable $J$ structures 
\begin{equation}
    \mathcal{M} = \{J\,|\, J \text{ is integrable}\}\quotient \GDiff_\bbC 
\end{equation}
Hence, to understand the local structure of the physical moduli space, we need to consider only deformations of $L_{-1}$ up to complex generalised diffeomorphisms.

Infinitesimally this can be reinterpreted as the cohomology of the following complex
\begin{equation}
\Gamma(E_{\mathbb{C}}) \xrightarrow{\; \dd_{1} \;} \Gamma(\adQ) \xrightarrow{\; \dd_{2} \;} \Gamma(W_{\mathbb{R}^{+}\times\Uni3\times\Spin{6+n}}^{\text{int}}), \label{eq:deformation complex}
\end{equation}
where $\adQ$ is a vector subbundle of $\ad\tilde{F}_{\mathbb{C}}$ such that $\Xi\cdot L_{-1} \nsubseteq L_{-1}$ for all non-zero sections $\Xi\in \Gamma(\adQ)$. We consider deformed bundles
\begin{equation}
L'_{-1} \coloneqq (1-\Xi)\cdot L_{-1} \qquad \Xi\in \Gamma(\adQ),
\end{equation}
such that the new $L'_{-1}$ is involutive with respect to the Dorfman derivative to linear order in $\Xi$. Since $L'_{-1}$ is involutive if and only if the intrinsic torsion of the corresponding $\mathbb{R}^{+}\times \Uni3\times \Spin{6+n}$ structure vanishes, this defines a linear map, denoted by $\dd_{2}$ above. The deformation is integrable if and only if $\Xi\in \kernel\dd_{2}$. There is also a notion of trivial deformations given by the action of complex generalised diffeomorphisms acting on $L_{-1}$. Infinitesimally this is just the Dorfman derivative along a complexified generalised vector. That is, a deformation $L'_{-1}=(1+\Xi)\cdot L_{-1}$ is trivial if there is some $V\in \Gamma(E_{\mathbb{C}})$ such that
\begin{equation}
L'_{-1} = (1+\Dorf_{V})L_{-1}.
\end{equation}
Again we can define a linear map $\dd_{1}$ such that a deformation generated by $\Xi\in \Gamma(\adQ)$ is trivial if and only if $\Xi\in \image\dd_{1}$. It is simple to show using \eqref{eq:dorfman commutator} that any trivial deformation is integrable and hence $\dd_{2}\circ \dd_{1}=0$. This means \eqref{eq:deformation complex} is a complex whose cohomology counts the physical moduli.

We will now find explicit expressions for the maps $\dd_{1}$ and $\dd_{2}$ using the parametrisation of $L_{-1}$ given in \eqref{eq:U(3)xSpin(6+n) L_-1}, and show that we recover the cohomology of \cite{OS14b,Ashmore:2018ybe}. Note that the choice of $\adQ$ is not unique for a given $L_{-1}$ and different choices change the form of the linear maps. A canonical choice comes from thinking of the fibres of $\adQ$ as quotient spaces $(\mathfrak{o}_{6,6+n}\oplus\mathbb{R})\quotient\mathfrak{p}$, where $\mathfrak{p}$ is the parabolic subalgebra preserving $L_{-1}$. Since we are only interested in the cohomology, which is independent of the exact choice of $\adQ$, we will choose convenient a representative.

Recall the form of $L_{-1}$
\begin{equation}
L_{-1} = \ee^{-B-\ii\,\omega}\ee^{-A}T^{0,1}.
\end{equation}
We take $\adQ$ to be
\begin{equation}
\adQ  \simeq \ee^{-B-\ii\,\omega}\ee^{-A}\cdot\bigl[(T^{1,0}\otimes T^{*0,1})\oplus\ext^{1,1}T^{*}_{\bbC}\oplus \ext^{0,2}T^{*} \oplus (T^{*0,1}\otimes\ad P_{G})\bigr]. \label{eq:choice of Q}
\end{equation}
We note that these bundles should be taken to be complexified as above, which we assume from this point forward. For any non-zero section $\Xi$ of this bundle we see that
\begin{equation}
\begin{split}
\Xi&\colon L_{-1} \to \ee^{-B-\ii\,\omega}\ee^{-A}(T^{1,0}\oplus T^{*}\oplus\ad P_{G}) \simeq E_{\mathbb{C}}\quotient L_{-1},\\
\Xi &= \ee^{-B-\ii\,\omega}\ee^{-A}\cdot(-\mu + x + b + \alpha)\in \Gamma(\adQ),
\end{split}
\end{equation}
where $\mu\in \Gamma(T^{1,0}\otimes T^{*0,1})$, $x\in \Gamma(T^{*1,1})$, $b\in \Gamma(T^{*0,2})$, and $\alpha\in \Gamma(T^{*0,1}\otimes\ad P_{G})$ -- these are what one might call complex structure, hermitian, and bundle moduli. (Again note that we are taking all of the bundles above to be complexified.) This shows that \eqref{eq:choice of Q} is a good choice of $\adQ$. We can then define our deformed bundle
\begin{equation}\label{eq:L1_deformed}
L'_{-1} = (1-\Xi)L_{-1}.
\end{equation}
To linear order in the deformation, we can rewrite this in a more convenient form as
\begin{equation}
L'_{-1} = \ee^{-\Theta}(1+\mu)T^{0,1},
\end{equation}
where $\Theta = B+\ii\omega +x+b+\tr(A\wedge\alpha) + A + \alpha$.\footnote{Here we note that to linear order $1+x+b+\alpha = \ee^{b+x}\ee^{\alpha}$ and then used the Baker--Campbell--Hausdorff formula together with \eqref{eq:AB_comm}.} It is worth stressing that by deforming within the space of $\UR$ structures we are including deformations that do not change the generalised metric, that is do not change the physical supergravity fields. In terms of the $\psi$ structure, the additional degrees of freedom parameterise $\Spin6/\SU3$ and transform in the $\rep{3}$ of $\SU3$ -- these correspond to deforming the putative Killing spinor, while keeping the supergravity fields fixed. If there are any such integrable deformations they would imply that the background actually defined an $\mathcal{N}=2$ rather than $\mathcal{N}=1$ solution. We will return to this point below.  

We now want to examine the conditions on $\Xi$ (or equivalently $\Theta$) for $L_{-1}'$ to be involutive, that is, for the deformation to be integrable. From \eqref{eq:L1_deformed}, two general sections $V,W\in \Gamma(L_{-1}')$ can be parametrised by $\Theta$, $\mu$ and two vectors $\bar v,\bar w\in\Gamma(T^{0,1})$. The Dorfman derivative of $W$ along $V$ can then be written in terms of a twisted derivative as
\begin{equation}
\begin{split}
\Dorf_{\ee^{-\Theta}(1+\mu)\bar v}\left(\ee^{-\Theta}(1+\mu)\bar w\right)
& =\ee^{-\Theta}\Dorf_{\bar v+\mu\cdot \bar v}^{\tilde{H}+\tilde{F}}(\bar w+\mu\cdot \bar w),
\end{split}
\end{equation}
where $\tilde{H}$ and $\tilde{F}$ are given to first order in the deformation by
\begin{align}
\begin{split}
\tilde{H} & =\dd B+\omega_{3}(A+\alpha)+\ii\,\dd\omega+\dd x+\dd b+\dd\tr(A\wedge\alpha)\\
& =2\ii\,\partial\omega+2\tr(\alpha\wedge F)+\dd x+\dd b,
\end{split}\\
\begin{split}
\tilde{F} & =\dd(A+\alpha)+(A+\alpha)\wedge(A+\alpha)\\
& =F+\dd_{A}\alpha,
\end{split}
\end{align}
where $\dd_A=\dd +[A,\cdot]$. Involutivity of $L_{-1}'$ is then equivalent to
\begin{equation}
\Dorf_{\bar v+\mu\cdot \bar v}^{\tilde{H}+\tilde{F}}(\bar w+\mu\cdot \bar w)=\bar u+\mu\cdot \bar u,
\end{equation}
for some $\bar u\in\Gamma(T^{0,1})$. Using the expression for the twisted Dorfman derivative from \eqref{eq:het_dorfman}, to first-order in the deformation we have
\begin{equation}
\begin{split}
\Dorf_{\bar v+\mu\cdot \bar v}^{\tilde{H}+\tilde{F}}(\bar w+\mu\cdot \bar w) & =[\bar v,\bar w]+[\mu\cdot \bar v,\bar w]+[\bar v,\mu\cdot \bar w]\\
& \eqspace-\imath_{\bar v}\imath_{\bar w}\bigl(2\tr(\alpha\wedge F)+\dd x+\dd b\bigr)-2\ii\,\imath_{\mu\cdot \bar v}\imath_{\bar w}\partial\omega-2\ii\,\imath_{\bar v}\imath_{\mu\cdot \bar w}\partial\omega\\
& \eqspace-\imath_{\bar v}\imath_{\bar w}\bar{\partial}_{A}\alpha-\imath_{\mu\cdot \bar v}\imath_{\bar w}F-\imath_{\bar v}\imath_{\mu\cdot \bar w}F\\
& \equiv \bar u+\mu\cdot \bar u.
\end{split}
\end{equation}
Decomposing according to complex type, we require
\begin{align}
[\bar v,\bar w]+[\mu\cdot \bar v,\bar w]^{0,1}+[\bar v,\mu\cdot \bar w]^{0,1} & =\bar u\label{eq:het_bundle_1}\\
[\mu\cdot \bar v,\bar w]^{1,0}+[\bar v,\mu\cdot \bar w]^{1,0} & =\mu\cdot \bar u,\label{eq:het_bundle_2}\\
\imath_{\bar v}\imath_{\bar w}\bar{\partial}_{A}\alpha+\imath_{\mu\cdot \bar v}\imath_{\bar w}F-\imath_{\mu\cdot \bar w}\imath_{\bar v}F & =0,\label{eq:het_bundle_3}\\
\imath_{\bar v}\imath_{\bar w}\bigl(2\tr(\alpha\wedge F)+\bar\partial x+\partial b\bigr)+2\ii\,\imath_{\mu\cdot \bar v}\imath_{\bar w}\partial\omega+2\ii\,\imath_{\bar v}\imath_{\mu\cdot \bar w}\partial\omega & =0,\label{eq:het_bundle_4}\\
\imath_{\bar v}\imath_{\bar w}\bar\partial b & =0.\label{eq:het_bundle_5}
\end{align}
Let us consider each of these conditions in turn. As we are working to first order in the deformations, dotting \eqref{eq:het_bundle_1} with $\mu$ and substituting into \eqref{eq:het_bundle_2} gives
\begin{equation}
\mu\cdot[\bar v,\bar w]=[\mu\cdot \bar v,\bar w]^{1,0}+[\bar v,\mu\cdot \bar w]^{1,0}.
\end{equation}
Expanding out in components and using a torsion-free compatible $\GL{3,\mathbb{C}}$ connection,\footnote{This exists as the undeformed solution admits an honest complex structure, $I$.} one can show this condition is equivalent to $\imath_{\bar w} \imath_{\bar v} \bar\partial\mu=0$, where $\mu$ is treated as a $(0,1)$-form with a holomorphic vector index. As this must vanish for all $\bar v$ and $\bar w$, we find
\begin{equation}
\bar\partial\mu=0.
\end{equation}
This is the expected condition on first-order deformations of a complex structure.

The third condition \eqref{eq:het_bundle_3} can be rewritten using $\imath_{\bar v}\imath_{\bar w}\imath_{\mu}F=\imath_{\mu\cdot \bar v}\imath_{\bar w}F-\imath_{\mu\cdot \bar w}\imath_{\bar v}F$, where $\imath_\mu F = e^a\wedge\imath_{\mu_a}F$, to give
\begin{equation}
\bar{\partial}_{A}\alpha+\imath_{\mu}F=0.
\end{equation}
The fourth condition \eqref{eq:het_bundle_4} can be rewritten using $\imath_{\bar v}\imath_{\mu\cdot \bar w}\partial\omega-\imath_{\bar w}\imath_{\mu\cdot \bar v}\partial\omega=-\imath_{\bar w}\imath_{\bar v}\imath_{\mu}\partial\omega$ to give
\begin{equation}
2\tr(\alpha\wedge F)+\bar{\partial} x+\partial b+2\ii\,\imath_{\mu}\partial\omega=0.
\end{equation}
The final condition \eqref{eq:het_bundle_5} is simply 
\begin{equation}
\bar{\partial}b=0.
\end{equation}

Taken together, the conditions are
\begin{align}
\bar{\partial}\mu & =0,\label{eq:het_moduli_1}\\
\bar{\partial}b & =0,\label{eq:het_moduli_2}\\
\bar{\partial}x+2\ii\,\imath_{\mu}\partial\omega+2\tr(\alpha\wedge F)+\partial b & =0,\label{eq:het_moduli_3}\\
\bar{\partial}_{A}\alpha+\imath_{\mu}F & =0\label{eq:het_moduli_4}.
\end{align}
These equations give the map $\dd_{2}$ on the different components of $\Xi$. It is comforting to note that these equations agree with those that have appeared before in work on heterotic moduli. To be precise, our equations match those in \cite{OS14b,Ashmore:2018ybe}, which we reproduce in \eqref{eq:AOMSCS 1} and \eqref{eq:AOMSCS 2}, after noting that $x_\text{here}=2 x_\text{there}$, $\mu_\text{here}=-\mu_\text{there}$ and $b_\text{here} = \mathcal{B}_\text{there}$.\footnote{The factor of two in $x$ is down to a choice of conventions. The minus sign that appears in $\mu$ is due to our $\mu$ deforming $T^{0,1}$ while the $\mu$ in \cite{OS14b,Ashmore:2018ybe} is a deformation of $T^{*1,0}$.} The only equation we are missing is \eqref{eq:AOMSCS 3} which is equivalent to the deformed complex three-form being conformally holomorphic. However, as we saw in section \ref{sec:moment}, this condition is imposed by the moment map, not involutivity. (Alternatively, one can see it as the extra condition that is imposed by the superpotential.) The particular missing equation is associated to the moment map condition that fixes $\psi$ (up to an overall constant) as a section of $\mathcal{U}_J$ once $J$ is determined. Since we have shown that we can describe the moduli space in terms of deformations of $J$ alone it does not appear. Note however, even if we had been using $\psi$ to parameterise the moduli space, we would still not have had to impose this relation. The point is that, as we have argued, at the level of the cohomology imposing moment map conditions is equivalent to quotienting by complex generalised diffeomorphisms. In other words, there will be representatives in the cohomology class for which this missing condition is satisfied and hence we do not need to impose it as an extra condition here. (This was actually the reason we could parameterise the moduli space using $J$ alone.) 
This illustrates the usefulness of this approach as it reduces the complexity of the equations governing the moduli. As a separate point, note also that the integrability conditions above are holomorphic in the complex parameters $\Xi$, as we would expect from our general discussion around \eqref{eq:inv_hol}.

We now examine the conditions for a deformation to be trivial. This will tell us what an ``exact'' deformation is and thus give the resulting cohomology that counts the inequivalent, non-trivial deformations. A deformation is to be regarded as trivial if the resulting $L_{-1}'$ is related to the undeformed subbundle by the action of the Dorfman derivative. In other words, if $L_{-1}'$ is simply a GDiff$_{\bbC}$ rotation of $L_{-1}$, the deformation is trivial. Let $V$ be a section of $L_{-1}$ and $W$ be a section of $E_\mathbb{C}$ such that
\begin{equation}
\begin{split}
V=e^{-B-\ii\,\omega}e^{-A}\bar{v},\\
W=e^{-B-\ii\,\omega}e^{-A}(w+\bar{w}+\xi+\bar{\xi}+\theta)=\ee^{-B-\ii\,\omega}\ee^{-A}W',
\end{split}
\end{equation}
where $w$ is a $(1,0)$-vector, $\bar v$ and $\bar w$ are $(0,1)$-vectors, $\xi$ and $\bar{\xi}$ are $(1,0)$- and $(0,1)$-forms, and $\theta$ is a complex gauge parameter. Note that $w$ and $\bar w$ (and $\xi$ and $\bar \xi$) are independent degrees of freedom and not related by complex conjugation, $\bar w \neq w^*$. Peeling off the twisting by $-B-\ii\,\omega$ and $-A$, the action of GDiff$_{\bbC}$ by $W$ on a section of $L_{-1}$ is 
\begin{equation}
\begin{split}\label{eq:dorf_trivial}
(1+\Dorf_{W'}^{H+\ii\,\dd\omega+F})\bar{v}	&=\bar{v}+[w+\bar{w},\bar{v}]-\imath_{\bar{v}}\dd(\xi+\bar{\xi})-\imath_{w+\bar{w}}\imath_{\bar{v}}(H+\ii\,\dd\omega)\\
&\eqspace+2\tr(\theta\,\imath_{\bar{v}}F)-\imath_{\bar{v}}\dd_{A}\theta-\imath_{w+\bar{w}}\imath_{\bar{v}}F\\
&=\bar{v}'-\imath_{\bar{v}'}\bar{\partial}w-\imath_{\bar{v}'}\bar{\partial}\xi-\imath_{\bar{v}'}\partial\bar{\xi}-\imath_{\bar{v}'}\bar{\partial}\bar{\xi}-2\ii\,\imath_{w}\imath_{\bar{v}'}\partial\omega\\
&\eqspace+2\tr(\theta\,\imath_{\bar{v}'}F)-\imath_{\bar{v}'}\bar{\partial}_{A}\theta-\imath_{w}\imath_{\bar{v}'}F,
\end{split}
\end{equation}
where $\bar{v}'=\bar{v}+[\bar{w},\bar{v}]+[w,\bar{v}]^{0,1}$ is a trivial rotation of $\bar{v}$ and we are working to first order in the components of $W$.

We want to compare this with the expression for a linear deformation of $L_{-1}$. Using the $\Orth{6,6+n}$ algebra~\cite{CMTW14} given in \eqref{eq:AB_comm} and the Baker--Campbell--Hausdorff formula, $L_{-1}'$ can be rewritten as
\begin{equation}
\begin{split}\label{eq:triv_def}
L_{-1}'&=\ee^{-B-\ii\,\omega-x-b-\tr(A\wedge\alpha)}\ee^{-A-\alpha}(1+\mu)\bar v\\
&=\ee^{-B-\ii\,\omega}\ee^{-A}(\bar{v}+\mu\cdot\bar{v}+\imath_{\bar{v}}x+\imath_{\bar{v}}b+\imath_{\bar{v}}\alpha).
\end{split}
\end{equation}
Comparing \eqref{eq:dorf_trivial} with the components in the parenthesis in \eqref{eq:triv_def}, one sees that a deformation of $L_{-1}'$ is actually the action of GDiff$_{\bbC}$, and so trivial, if
\begin{align}
\mu&=-\bar{\partial}w, \label{eq:het_exact_1}\\
x&=-\bar{\partial}\xi-\partial\bar{\xi}+2\ii\,\imath_{w}\partial\omega+2\tr(\theta\,F), \label{eq:het_exact_2}\\
b&=-\bar{\partial}\bar{\xi}, \label{eq:het_exact_3}\\
\alpha&=-\bar{\partial}_{A}\theta+\imath_{w}F. \label{eq:het_exact_4}
\end{align}
Combined, these derivatives form the operator $\dd_{1}$. One can check that these satisfy \eqref{eq:het_moduli_1}--\eqref{eq:het_moduli_4} (so that exact deformations are automatically closed) provided $\{\partial,\bar\partial\}=0$, $\bar\partial^2=\bar\partial_A^2=0$, implying the original solution has a complex structure and a holomorphic gauge bundle, and $F$ and $H$ satisfy the appropriate Bianchi identities. These will each hold as we are assuming we are deforming around an $\mathcal{N}=1$ solution. Combining \eqref{eq:het_exact_1}--\eqref{eq:het_exact_4} with \eqref{eq:het_moduli_1}--\eqref{eq:het_moduli_4}, we recover precisely the cohomology of \cite{OS14b} up to the $b$ term which is not present in their analysis. This is included in the linear terms in the same calculation in \cite{Ashmore:2018ybe} and is related to deformations of $B_{0,2}$.


It is worth analysing this $b$ modulus further. 
As we mentioned above, our parameterisation of the deformation includes not only deformations of the physical fields preserving $\mathcal{N}=1$ supersymmetry but also potential deformations of the Killing spinors, with the same background geometry. The latter type of deformations correspond to the background admitting additional supersymmetries. Specifically one can show that a particular combination of $b$ and $\mu$ will leave the generalised metric invariant and hence correspond to such additional supersymmetries. From the form of the equations~\eqref{eq:het_moduli_2} and~\eqref{eq:het_exact_2} we see that if $h^{0,2}$ vanishes then there are no moduli for deformations of $b$ and hence all the deformations correspond to physical deformations of the background -- in other words this is sufficient for the background not to admit additional supersymmetries. A counter example is the solution on $\text{K3}\times \text{T}^{2}$ with trivial gauge group. In this case $h^{0,2}\neq0$ and the $b$ modulus survives. The additional degree of freedom corresponds to rotating the choice of $\mathcal{N}=1$ subalgebra picked out by $\psi$ within the $\mathcal{N}=2$ supersymmetry algebra.

\section{Conclusions}\label{sec:conclusions}

We have shown that the Hull--Strominger system can be reformulated as an integrable $\SU3\times\Spin{6+n}$ structure within $\Orth{6,6+n}\times\mathbb{R}^{+}$ generalised geometry. The structure is defined by a particular generalised tensor $\psi\in\Gamma(\ext^{3}E\otimes\ext^{6}T^{*})$, where supersymmetry for the background is equivalent to the differential condition that the structure is torsion-free or integrable. The integrability conditions for $\psi$ split into an involutivity condition of a subbundle $L_{-1}\subset E$ (the ``F-term'' condition) and the vanishing of a moment map for the action of generalised diffeomorphisms on the space of structures (the ``D-term'' condition). Furthermore, this formalism gives $\Orth{6,6+n}\times\mathbb{R}^{+}$ covariant expressions for both the superpotential and Kähler potential of a generic off-shell heterotic background. 

Starting with the work of Fu, Li and Yau~\cite{li2005,Fu:2006vj}, several constructions  of explicit solutions to the Hull--Strominger system are now known (for a review see~\cite{GF16}). It would, of course, be interesting to have theorems about the existence and uniqueness of solutions, and some steps in this direction were made in the work of~\cite{GRST18} which showed that the system could be reinterpreted in terms of an extremisation problem within a particular class of holomorphic Courant algebroids. Our work gives a reinterpretation of this structure, in particular showing that it follows from a moment map, as for the conventional Calabi--Yau case. Specifically, the holomorphic algebroid is determined by the solution of the F-terms, that is the choice of involutive sub-bundle $L_{-1}$. Because the space of such structures is Kähler, solving the moment map is equivalent to extremizing the Kähler potential, which is indeed the ``dilaton functional'' discussed in~\cite{GRST18}. We discussed briefly how this setup defines a GIT quotient that includes as special cases both the standard notion of stability for the hermitian Yang--Mills equations (that is, for the gauge fields, suitably generalised to non-Kähler backgrounds) and the notion of stability of Calabi--Yau metrics.


We also studied the moduli of these backgrounds by reformulating the problem in terms of finding integrable deformations of the subbundle $L_{-1}$ up to complexified generalised diffeomorphisms. From this we were able to match to the known $\bar{D}$ cohomology of \cite{OS14b,Ashmore:2018ybe} with considerably less work. In doing so, we defined the differentials $\dd_1$ and $\dd_2$ that appear in the relevant complex. Note however that there is another natural differential associated with the structures. The subbundle $L_{-1}$ is involutive and, since $L_{-1}$ is isotropic, $\eta(L_{-1},L_{-1})=0$, the Dorfman derivative satisfies a Jacobi identity (while generic sections of $E$ do not). Together, this means that $L_{-1}$ defines a Lie algebroid and hence comes with a natural differential $\dd_{L}$. Following \cite{Gualtieri:2017kdd}, it would be interesting to see how this relates to the differential $\bar{D}$ found in \cite{OS14b} and whether the cohomology that counts the moduli can be reformulated in terms of $\dd_{L}$. We hope to return to this in the future.

It is natural to ask whether we can use our formalism to explore finite deformations of the background and compare this to the results found in \cite{Ashmore:2018ybe}. Note that this would require understanding whether the deformations are obstructed. For the infinitesimal moduli, the moment map condition is imposed indirectly via the quotient by $\GDiff_{\mathbb{C}}$. For this to work, there must be some deformed $\psi\in\hat{\mathcal{Z}}$ in the orbit of $\GDiff_{\mathbb{C}}$ that satisfies the moment map constraint. If the moment map is well behaved and the $\GDiff$ action has no fixed points, the moment map implies that solutions will always exist in some finite neighbourhood, that is there will be no obstructions when we go beyond first order (though there may be some ``jumping'' behaviour when the deformation gets large enough). In our set-up, fixed points correspond to a supersymmetric background which is invariant under some action of  $L_V\in\gdiff$ that also preserves $\psi$. Since the generalised metric (and hence the conventional metric) are determined by $\psi$, these must be isometries that preserve the full solution. For the case where the original solution is simply Calabi--Yau (viewed as an $\mathcal{N}=1$ solution), there are no continuous symmetries that preserve the solution and hence we would argue that the deformations are unobstructed. Note that this goes beyond the usual statement that the Calabi--Yau moduli space is unobstructed, as this also includes turning on flux and deforming the gauge bundle.

We might also ask how much of this structure is relevant to other string backgrounds. First we note that backgrounds with higher supersymemtry can always be viewed as $\mathcal{N}=1$ solutions. Thus, for example, by choosing a particular $\mathcal{N}=1$ subalgebra, our calculation is good for both the $\mathcal{N}=1$ and $\mathcal{N}=2$ solutions of~\cite{Dasgupta:1999ss,BS09,Melnikov:2014ywa}. In the latter case, following the discussion at the end of the last section, we would expect to find additional moduli corresponding to deforming the choice of $\mathcal{N}=1$ inside $\mathcal{N}=2$.
More broadly, generic backgrounds in type II or M-theory that preserve eight supercharges can also be described in generalised geometry~\cite{AW15}, where now one has two compatible structures (dubbed V and H structures). The H structure is defined by an $\SU2$ triplet of generalised tensors $J_\alpha$ transforming in the adjoint bundle. These define a $G$-structure where the particular group depends on the dimension of the external space. Similar to our discussion in this paper, one can pick out a weaker $\mathbb{C}^{*}\times G$ structure, defined by $J_+$ alone. Again, this turns out to be defined by a subbundle of the generalised tangent, with the corresponding integrability conditions coming from involutivity. In the AdS case~\cite{AW15b}, since deformations of the structure are ``holomorphic'' in a certain sense, this formulation might provide a way to explore the conformal manifold of the dual CFTs. We hope to make progress on this in the near future.

In a similar vein, an obvious application of the analysis used here and in~\cite{N1paper} is to $\AdS4$ backgrounds in M-theory. Previous work on backgrounds which preserve eight supersymmetries showed that generalised geometry could be used to understand properties of the dual three-dimensional CFTs with $\mathcal{N}=2$ supersymmetry~\cite{AGGPW16,Ashmore:2018npi}. It would be interesting to use the $\mathcal{N}=1$ language developed in this paper and \cite{N1paper} to extend this analysis to $\mathcal{N}=1$ CFTs in three dimensions. Unlike the Kähler structure on the moduli space that we encountered in this work, we expect the moduli space to have a real structure. While the moduli space itself will again come from an involutivity condition and a moment map, we expect that this will not have a picture as a complexified quotient, but rather simply be a symplectic quotient. It would be interesting to identify the corresponding picture in the dual field theory. A final direction would be to try to match our description of the moduli space of heterotic compactifications to that of ``universal geometry'', as has appeared in \cite{COM17,Candelas:2018lib,McOrist:2019mxh}. There one finds that the resulting moduli space is beautifully described by combining the heterotic geometry and parameter space into a single space, with the geometry fibred over the parameter space. This allows one to write differential operators that act on the total space, leading to compact expressions for both the linear deformation conditions and the Kähler potential of the background.

\acknowledgments

We thank Ed Tasker, Ruben Minasian, Eirik Eik Svanes and Xenia de la Ossa for helpful discussions. AA is supported by DOE Grant No.~DESC0007901. DT is supported by an STFC PhD studentship.  DW is supported in part by the STFC Consolidated Grant ST/P000762/1. We acknowledge the Mainz Institute for Theoretical Physics (MITP) of the Cluster of Excellence PRISMA+ (Project ID 39083149) and the Berkeley Center for Theoretical Physics at UC Berkeley for hospitality and support during part of this work. 

\appendix

\section{\texorpdfstring{$\Orth{6,6+n}$}{O(6,6+n)} generalised geometry}\label{ap:O(6,6+n) generalised geometry}

Here we collect a number of useful formula for the $\Orth{6,6+n}\times\mathbb{R}^{+}$ generalised geometry relevant for type I and heterotic backgrounds. A more detailed discussion can be found in \cite{CMTW14}.

The adjoint action of a two-form $B$, a two-vector $\beta$ and a one-form gauge field $A$ on a generalised vector $V=v+\lambda+\Lambda$ are given by
\begin{equation}
\begin{split}
\ee^{B}V&=v+\lambda-\imath_{v}B+\Lambda,\\
\ee^{\beta}V&=v+\lambda-\beta\lrcorner\lambda+\Lambda,\\
\ee^{A}V&=v+\lambda+2\tr(\Lambda A)-\tr(\imath_{v}A\,A)+\Lambda-\imath_{v}A.
\end{split}
\end{equation}
Note that $B$ commutes with itself, while $A$ has a non-trivial commutator with itself
\begin{equation}\label{eq:AB_comm}
[A,A']=-2\tr(A\wedge A').
\end{equation}
One can check that the natural inner product
\begin{equation}
\eta(v+\Lambda + \lambda,w + \Sigma +\sigma) = \tfrac{1}{2}\imath_{v}\sigma + \tfrac{1}{2}\imath_{w}\lambda +\tr(\Lambda\Sigma),
\end{equation}
is preserved by the above action.

The twisted Dorfman derivative is defined by
\begin{equation}
\Dorf_{\ee^{-B}\ee^{-A}V}(\ee^{-B}\ee^{-A}W)=\ee^{-B}\ee^{-A}\Dorf_{V}^{H+F}W,
\end{equation}
where for $V=v+\lambda+\Lambda$ and $W=w+\rho+\Sigma$, we have
\begin{align}
\Dorf_{V}^{H+F}W & =[v,w]\nonumber\\
& \eqspace+\mathcal{L}_{v}\rho-\imath_{w}\dd\lambda-\imath_{v}\imath_{w}H+2\tr(\Sigma\,\dd_{A}\Lambda)-2\tr(\Sigma\,\imath_{v}F)+2\tr(\Lambda\,\imath_{w}F)\label{eq:het_dorfman}\\
& \eqspace+[\Lambda,\Sigma]+\imath_{v}\dd_{A}\Sigma-\imath_{w}\dd_{A}\Lambda-\imath_{v}\imath_{w}F,\nonumber
\end{align}
where we have defined
\begin{align}
\dd_{A}\Lambda & =\dd\Lambda+[A,\Lambda],\\
F & =\dd A+A\wedge A,\\
H & =\dd B+\tr(A\wedge\dd A+\tfrac{2}{3}A\wedge A\wedge A).
\end{align}
We also have the usual rule for the commutator of two Dorfman derivatives
\begin{equation}
[L_{U},L_{V}]W=L_{L_{U}V}W=\Dorf_{\llbracket U,V\rrbracket} W, \label{eq:dorfman commutator}
\end{equation}
where $\llbracket\cdot,\cdot\rrbracket$ is the Courant bracket, the antisymmetrisation of the Dorfman derivative.

\section{Explicit calculations of the superpotential, Kähler potential and moment map}\label{sec:explicit_moment}

In this appendix, we lay out in detail how one calculates the superpotential, Kähler potential and the moment map using the explicit form of $\psi$ and $J$ given in the main text.

\subsection{The superpotential}

To see that our expression for the superpotential~\eqref{eq:superpotential} matches the conventional expression given in~\eqref{eq:het_superpotential}, we expand in $\Orth{6,6+n}$ indices:
\begin{equation}
\begin{split}
W &\sim \int_{X} J^{A}{}_{B}\Dgen_{C}\psi^{CB}{}_{A} \\
&\sim \int_{X}\Dgen_{C}(J^{A}{}_{B}\psi^{CB}{}_{A}) - \psi^{CBA} \Dgen_{\left[C\right. }J_{\left.AB \right]} \\
&\sim \int_{X}\psi^{ABC}\Dgen_{\left[A\right.} J_{\left. BC\right]} \\
&\sim \int_{X}\sqrt{g}\,\ee^{-2\varphi}\Omega^{\bar{\mu}\bar{\nu}\bar{\rho}} \Dgen_{\left[\bar{\mu}\right.} J_{\left. \bar{\nu}\bar{\rho}\right]},
\end{split}
\end{equation}
where we have used the fact that the boundary term vanishes identically, and have raised/lowered indices with $\eta$. To reach the final lines we have used results from the previous section on the contraction of $\psi$ with a section of $\ext^{3}E$. Hence all that remains is to determine the form of $ \Dgen_{\left[\bar{\mu}\right.} J_{\left. \bar{\nu}\bar{\rho}\right]}$. Using the components of the connection from~\cite{Garcia-Fernandez:2013gja}, we have that
\begin{equation}
\begin{split}
\Dgen_{[\bar{\mu}} J_{\bar{\nu}\bar{\rho}]} &= \nabla_{[\bar{\mu}} J_{ \bar{\nu}\bar{\rho}]} - \tfrac{1}{3}H_{[\bar{\mu}}{} ^{\sigma}{}_{\bar{\nu}|}J_{|\sigma| \bar{\rho}]} \\
&= \tfrac{1}{3}(-\dd\omega)_{\bar{\mu}\bar{\nu}\bar{\rho}} + \tfrac{\ii}{3}H_{[\bar{\mu}}{} ^{\sigma}{}_{\bar{\nu}}g_{|\sigma|\bar{\rho}]} \\
&\sim (H+\ii \dd\omega)_{\bar{\mu}\bar{\nu}\bar{\rho}},
\end{split}
\end{equation}
where we have used $g_{\mu\bar{\nu}}=-\ii\omega_{\mu\bar{\nu}}$ for an $\SU3$ structure. Hence
\begin{equation}
\begin{split}
\mathcal{W} &\sim \int_{X} \sqrt{g}\,\ee^{-2\varphi}\Omega^{\bar{\mu}\bar{\nu}\bar{\rho}}(H+\ii\, \dd\omega)_{\bar{\mu}\bar{\nu}\bar{\rho}} \\
&\sim \int_{X}\ee^{-2\varphi}\Omega\wedge(H+\ii\,\dd\omega).
\end{split}
\end{equation}
This is precisely the form of the superpotential in \eqref{eq:het_superpotential} and used in \cite{OHS16,McOrist16}. Hence our expression \eqref{eq:superpotential} is the covariant form of the superpotential for a generic four-dimensional $\mathcal{N}=1$ heterotic background determined by $\psi$.

\subsection{The Kähler potential}

The Kähler potential is
\begin{equation}
    \mathcal{K}=\int_X \eta(\psi,\bar\psi)^{1/2},
\end{equation}
where $\eta$ is the symmetric pairing on sections of $\ext^{3}E$. We fix our conventions for this in terms of $\eta$ on sections of $E$ by examining how the usual inner product defined by $g$ acts on tri-vectors. For $\alpha,\beta\in\Gamma(\ext^{3}T)$, the pairing is
\begin{equation}
\begin{split}
g(\alpha,\beta) & =\tfrac{1}{3!}\tfrac{1}{3!}\alpha^{mnp}\beta^{qrs}g(\hat{e}_{mnp},\hat{e}_{qrs})\\
& \equiv\tfrac{1}{3!}\alpha^{mnp}\beta^{qrs}g(\hat{e}_{m},\hat{e}_{q})g(\hat{e}_{n},\hat{e}_{r})g(\hat{e}_{p},\hat{e}_{s})\\
& =\tfrac{1}{3!}\alpha^{mnp}\beta_{mnp}\\
& =\alpha\lrcorner\beta,
\end{split}
\end{equation}
where we have used $\hat{e}_{mnp}\lrcorner e^{qrs}=3!\delta_{[m}^{q}\delta_{n}^{r}\delta_{p]}^{s}$. Similarly we define
\begin{equation}
\begin{split}
\eta(\hat{E}_{mnp}^{+},\hat{E}_{qrs}^{+}) & =3!\eta(\hat{E}_{m}^{+},\hat{E}_{q}^{+})\eta(\hat{E}_{n}^{+},\hat{E}_{r}^{+})\eta(\hat{E}_{p}^{+},\hat{E}_{s}^{+})\\
& =3!\delta_{mq}\delta_{nr}\delta_{ps},
\end{split}
\end{equation}
where an antisymmetrisation over $mnp$ is assumed and for simplicity we take $\hat e_{m}$ to be an orthonormal frame, implying $\eta(\hat{E}_{m}^{+},\hat{E}_{n}^{+})=g_{mn}=\delta_{mn}$. With $\chi$ defined as in \eqref{eq:chi_def}
\begin{equation}
\chi = \frac{1}{3!}g^{1/4}\ee^{-\varphi}\Omega^{mnp} \hat{E}^{+}_{mnp},
\end{equation}
the pairing $\eta(\chi,\bar{\chi})$ is given by
\begin{equation}
\begin{split}
\eta(\chi,\bar{\chi}) 
& =\tfrac{1}{3!}g^{1/2}\ee^{-2\varphi}\Omega^{mnp}\bar{\Omega}_{mnp}\\
& =g^{1/2}\ee^{-2\varphi}\Omega^{\sharp}\lrcorner\bar{\Omega}\\
& =\ii\,\ee^{-2\varphi}\Omega\wedge\bar{\Omega},
\end{split}
\end{equation}
where we have used the standard $\SU 3$ structure relations
\begin{equation}
\Omega^{\sharp}\lrcorner\bar{\Omega}=8,\qquad g^{1/2}=\vol=\frac{\ii}{8}\Omega\wedge\bar{\Omega}.
\end{equation}
Integrated over $X$, this gives the expression for the Kähler potential given in the main text.

\subsection{The moment map}

The expression for the moment map given in the main text is
\begin{equation}
\mu(V)=-\frac{\ii}{2}\int_X\eta(\Dorf_{V}\chi,\bar{\chi}).
\end{equation}
To evaluate this, we need an expression for the Dorfman derivative of $\chi$. For $V=\ee^{-B}\ee^{-A}(v+\lambda+\Lambda)$, where $v\in\Gamma(T)$, $\lambda\in\Gamma(T^{*})$ and $\Lambda\in\Gamma(\ad P)$, we have
\begin{align}
\Dorf_{V}\chi & =\tfrac{1}{3!}\mathcal{L}_{v}(g^{1/4}\ee^{-\varphi}\Omega^{mnp})\hat{E}_{mnp}^{+}+\tfrac{1}{2}g^{1/4}\ee^{-\varphi}\Omega^{mnp}\Dorf_{V}\hat{E}_{m}^{+}\wedge\hat{E}_{np}^{+},\\
\Dorf_{V}\hat{E}_{m}^{+} & =\ee^{-B}\ee^{-A}\left(\mathcal{L}_{v}(\hat{e}_{m}+e_{m})-\imath_{\hat{e}_{m}}\dd\lambda-\imath_{v}\imath_{\hat{e}_{m}}H+2\tr(\Lambda\imath_{\hat{e}_{m}}F)-\imath_{\hat{e}_{m}}\dd_{A}\Lambda-\imath_{v}\imath_{\hat{e}_{m}}F\right).
\end{align}
The expression for the moment map is then
\begin{equation}
\begin{split}
\mu(V) 
& =-\tfrac{\ii}{2}\int_X\eta\left(\tfrac{1}{3!}\mathcal{L}_{v}(g^{1/4}\ee^{-\varphi}\Omega^{mnp})\hat{E}_{mnp}^{+}+\tfrac{1}{2}g^{1/4}\ee^{-\varphi}\Omega^{mnp}\Dorf_{V}\hat{E}_{m}^{+}\wedge\hat{E}_{np}^{+},\tfrac{1}{3!}g^{1/4}\ee^{-\varphi}\bar{\Omega}^{qrs}\hat{E}_{qrs}^{+}\right)\\
& =-\tfrac{\ii}{2}\int_X\tfrac{1}{3!}\mathcal{L}_{v}(g^{1/4}\ee^{-\varphi}\Omega^{mnp})g^{1/4}\ee^{-\varphi}\bar{\Omega}_{mnp}\\
& \eqspace-\tfrac{\ii}{2}\int_X\tfrac{1}{2}g^{1/4}\ee^{-\varphi}\Omega^{mnp}\tfrac{1}{3!}g^{1/4}\ee^{-\varphi}\bar{\Omega}^{qrs}\eta(\Dorf_{V}\hat{E}_{m}^{+}\wedge\hat{E}_{np}^{+},\hat{E}_{qrs}^{+}),
\end{split}
\end{equation}
where we have used
\begin{equation}
\eta(\Dorf_{V}\hat{E}_{m}^{+}\wedge\hat{E}_{np}^{+},\hat{E}_{qrs}^{+})  
=3!\,\eta(\Dorf_{V}\hat{E}_{m}^{+},\hat{E}_{q}^{+})\delta_{nr}\delta_{ps},
\end{equation}
with an assumed antisymmetrisation over $mnp$ and
\begin{equation}
\begin{split}
\eta(\Dorf_{V}\hat{E}_{m}^{+},\hat{E}_{n}^{+}) & =\eta\left(\mathcal{L}_{v}(\hat{e}_{m}+e_{m})-\imath_{\hat{e}_{m}}\dd\lambda-\imath_{v}\imath_{\hat{e}_{m}}H+2\tr(\Lambda\imath_{\hat{e}_{m}}F)-\imath_{\hat{e}_{m}}\dd_{A}\Lambda-\imath_{v}\imath_{\hat{e}_{m}}F,\hat{e}_{n}+e_{n}\right)\\
& =\tfrac{1}{2}\imath_{\mathcal{L}_{v}\hat{e}_{m}}e_{n}+\tfrac{1}{2}\imath_{\hat{e}_{n}}\mathcal{L}_{v}e_{m}-\tfrac{1}{2}\imath_{\hat{e}_{n}}\imath_{\hat{e}_{m}}\dd\lambda-\tfrac{1}{2}\imath_{\hat{e}_{n}}\imath_{v}\imath_{\hat{e}_{m}}H+\imath_{\hat{e}_{n}}\tr(\Lambda\imath_{\hat{e}_{m}}F).
\end{split}
\end{equation}
Our task is now to find what conditions $\mu=0$ imposes. To do this, we examine $\mu(V)=0$ where $V$ consists of an arbitrary vector, one-form or gauge parameter in turn. First, consider $V=\ee^{-B}\ee^{-A}\lambda$:
\begin{equation}\label{eq:mu_1}
\begin{split}
\int_X\eta(\Dorf_{V}\chi,\bar{\chi})
& =\int_X\tfrac{1}{2}g^{1/4}\ee^{-\varphi}\Omega^{mnp}g^{1/4}\ee^{-\varphi}\bar{\Omega}^{qrs}\left(-\tfrac{1}{2}\right)\imath_{\hat{e}_{q}}\imath_{\hat{e}_{m}}\dd\lambda\,\delta_{nr}\delta_{ps}\\
& =-\tfrac{1}{4}\int_X\ee^{-2\varphi}\Omega^{mnp}\bar{\Omega}^{q}{}_{np}\imath_{\hat{e}_{q}}\imath_{\hat{e}_{m}}\dd\lambda\,\vol\\
& =2\ii\int_X\ee^{-2\varphi}\dd\lambda\wedge\omega\wedge\omega\\
& =2\ii\int_X\lambda\wedge\dd(\ee^{-2\varphi}\omega\wedge\omega),
\end{split}
\end{equation}
where we have used the $\SU3$ structure identity
\begin{equation}\label{eq:random_identity}
\Omega^{mnp}\bar{\Omega}^{q}{}_{np}(\imath_{\hat{e}_{q}}\imath_{\hat{e}_{m}}\alpha_{2})\,\vol=-8\ii\alpha_{2}\wedge\omega\wedge\omega,
\end{equation}
which holds for an arbitrary two-form $\alpha_{2}$.

Next, consider $V=\ee^{-B}\ee^{-A}\Lambda$:
\begin{equation}\label{eq:mu_2}
\begin{split}
\int_X\eta(\Dorf_{V}\chi,\bar{\chi})& =\int_X\tfrac{1}{2}g^{1/4}\ee^{-\varphi}\Omega^{mnp}\tfrac{1}{3!}g^{1/4}\ee^{-\varphi}\bar{\Omega}^{qrs}3!\left(\imath_{\hat{e}_{q}}\tr(\Lambda\imath_{\hat{e}_{m}}F)\right)\delta_{nr}\delta_{ps}\\
& =\int_X\tfrac{1}{2}\vol\,\ee^{-2\varphi}\Omega^{mnp}\bar{\Omega}^{q}{}_{np}\imath_{\hat{e}_{q}}\imath_{\hat{e}_{m}}\tr(\Lambda F)\\
& =\int_X\tfrac{1}{2}\ee^{-2\varphi}(-8\ii)\,\tr(\Lambda F)\wedge\omega\wedge\omega\\
& =-4\ii\int_X\tr(\Lambda F)\wedge\ee^{-2\varphi}\omega\wedge\omega,
\end{split}
\end{equation}
where again we have used \eqref{eq:random_identity}.

Finally, consider $V=\ee^{-B}\ee^{-A}v$:
\begin{equation}
\begin{split}
\int_X\eta(\Dorf_{V}\chi,\bar{\chi})
& =\int_X\tfrac{1}{3!}\mathcal{L}_{v}(g^{1/4})g^{1/4}\ee^{-2\varphi}8\cdot3!+\tfrac{1}{3!}\mathcal{L}_{v}(\ee^{-\varphi}\Omega^{mnp})g^{1/2}\ee^{-\varphi}\bar{\Omega}_{mnp}\\
& \eqspace+\int_X\tfrac{1}{4}g^{1/2}\ee^{-2\varphi}\Omega^{mnp}\bar{\Omega}^{q}{}_{np}\left(\imath_{\hat{e}_{q}}\mathcal{L}_{v}e_{m}-\imath_{\hat{e}_{m}}\mathcal{L}_{v}e_{q}-\imath_{\hat{e}_{q}}\imath_{v}\imath_{\hat{e}_{m}}H\right).
\end{split}
\end{equation}
Now note that the first term is real while $\int_X\eta(\Dorf_{V}\chi,\bar{\chi})$ is imaginary (after an integration by parts), so it cancels. The remaining terms can be rewritten as
\begin{equation}\label{eq:moment_intermediate}
\begin{split}
\int_X\eta(\Dorf_{V}\chi,\bar{\chi})
& =\int_X\tfrac{1}{2}\tfrac{1}{3!}g^{1/2}\ee^{-2\varphi}\mathcal{L}_{v}\Omega^{mnp}\bar{\Omega}_{mnp}+\tfrac{1}{2}\tfrac{1}{3!}g^{1/2}\ee^{-2\varphi}\mathcal{L}_{v}\bar{\Omega}^{mnp}\Omega_{mnp}\\
& \eqspace+\int_X\tfrac{1}{4}g^{1/2}\ee^{-2\varphi}\left(2\bar{\Omega}\lrcorner\mathcal{L}_{v}\Omega-2\Omega\lrcorner\mathcal{L}_{v}\bar{\Omega}-\Omega^{mnp}\bar{\Omega}^{q}{}_{np}\imath_{\hat{e}_{q}}\imath_{v}\imath_{\hat{e}_{m}}H\right),
\end{split}
\end{equation}
where we have the $\SU3$ structure identities $\Omega^{\sharp}\lrcorner\bar{\Omega} =8$ and $8\vol=\ii\Omega\wedge\bar{\Omega}$, and
\begin{align}
\Omega^{mnp}\bar{\Omega}^{q}{}_{np} & =8g^{mq}+8\ii\,I^{mq}=8g^{mq}-8\ii\,\omega^{mq},\\
2\bar{\Omega}\lrcorner\mathcal{L}_{v}\Omega & =\tfrac{1}{3}\mathcal{L}_{v}\Omega^{mnp}\bar{\Omega}_{mnp}+\Omega^{mnp}\bar{\Omega}^{q}{}_{np}\imath_{\hat{e}_{q}}\mathcal{L}_{v}e_{m},\\
2\Omega\lrcorner\mathcal{L}_{v}\bar{\Omega} & =\tfrac{1}{3}\mathcal{L}_{v}\bar{\Omega}^{mnp}\Omega_{mnp}+\Omega^{mnp}\bar{\Omega}^{q}{}_{np}\imath_{\hat{e}_{m}}\mathcal{L}_{v}e_{q}.
\end{align}
Again, note that the first two terms of \eqref{eq:moment_intermediate} combine to give something real, and so they must cancel. We can then massage the remaining terms to give
\begin{equation}\label{eq:mu_3}
\begin{split}
\int_X\eta(\Dorf_{V}\chi,\bar{\chi})
& =\int_X\tfrac{1}{4}\ee^{-2\varphi}\left(2\ii\mathcal{L}_{v}\Omega\wedge\bar{\Omega}+2\ii\mathcal{L}_{v}\bar{\Omega}\wedge\Omega-8\ii\imath_{v}H\wedge\omega\wedge\omega\right)\\
& =\ii\tfrac{1}{2}\int_X\ee^{-2\varphi}(2\imath_{v}\bar{a}-2\imath_{v}a+2\imath_{v}\partial\varphi-2\imath_{v}\bar{\partial}\varphi)\,\Omega\wedge\bar{\Omega}\\
& \eqspace+2\ee^{-2\varphi}\imath_{v}\partial\varphi\,\Omega\wedge\bar{\Omega}-2\ee^{-2\varphi}\imath_{v}\bar{\partial}\varphi\Omega\wedge\bar{\Omega}\\
& =\ii\int_X\ee^{-2\varphi}(\imath_{v}\bar{a}-\imath_{v}a+2\imath_{v}\partial\varphi-2\imath_{v}\bar{\partial}\varphi)\,\Omega\wedge\bar{\Omega}.
\end{split}
\end{equation}
To reach this result, we have integrated by parts and used $\dd\Omega=\bar{a}\wedge\Omega$ for $\bar{a}\in \Omega^{0,1}(X)$, which is implied by integrability of the complex structure which in turn comes from involutivity of $L_{-1}$. We have also used $\bar{\Omega}\lrcorner\alpha_{3} \vol  =\ii\,\alpha_{3}\wedge\bar{\Omega}$ for an arbitrary three-form $\alpha_{3}$. Summed up, the three contributions to $\mu(V)$ in \eqref{eq:mu_1}, \eqref{eq:mu_2} and \eqref{eq:mu_3} give the expression for the moment map given in the main text.

\input{main.bbl}
\end{document}

%% file: main.bbl
\providecommand{\href}[2]{#2}\begingroup\raggedright\endgroup